\documentclass[12pt]{article} 

\usepackage[T1]{fontenc}
\usepackage[tracking=true]{microtype}
\usepackage{lmodern}
\usepackage{verbatim}
\usepackage{amssymb}

\usepackage[margin=1in]{geometry}

\usepackage{verbatim}
\usepackage{graphicx}
\usepackage{subcaption}
\usepackage{booktabs}
\usepackage{longtable}
\usepackage{cite}
\usepackage{amsmath}
\usepackage{empheq}

\usepackage{color}
\definecolor{darkblue}{cmyk}{0.9,0.9,0,0}
\definecolor{wine-stain}{rgb}{0.5,0,0}
\usepackage[colorlinks, allcolors=darkblue, linktoc=all]{hyperref} 
\usepackage{setspace}

\renewcommand{\L}{\mathcal L}

\newcommand{\beq}{\begin{equation}}
\newcommand{\eeq}{\end{equation}}
\newcommand{\be}{\begin{equation}}
\newcommand{\ee}{\end{equation}}
\newcommand{\bea}{\begin{align}}
\newcommand{\eea}{\end{align}}

\def\gsim{ \lower .75ex \hbox{$\sim$} \llap{\raise .27ex \hbox{$>$}} }
\def\lsim{ \lower .75ex \hbox{$\sim$} \llap{\raise .27ex \hbox{$<$}} }

\def\beq{\begin{eqnarray}}
\def\eeq{\end{eqnarray}}

\def\mpl{M_{\rm Pl}}

\def\p{{\cal P}}

\def\L*{{\cal L}_*}
\def\L{\mathcal{L}}
\def\({\left(}
\def\){\right)}

\def\nn{\nonumber}
\def\p{\partial}

\def\stu{St\"uckelberg }
\def\p{\partial}

\def\<{\langle}
\def\>{\rangle}

\def\xyma{\xymatrix@M.7em}
\def\xymas{\xymatrix@M.1em}
\newcommand{\ba}{\begin{eqnarray}}
\newcommand{\ea}{\end{eqnarray}}

\begin{document}
\thispagestyle{empty}
\vspace*{0.5in} 
\begin{center}
{\Huge {Resolving the vDVZ and Strong Coupling \\ \vskip 0.3cm Problems in Massive Gravity and Bigravity}}

\vskip 0.7cm
\centerline{
{\large {Gregory Gabadadze,}}
{\large {Daniel Older,}}
{\large {and David Pirtskhalava}}
}
\vspace{.2in} 

\centerline{{\it Center for Cosmology and Particle Physics, Department of Physics}}
\centerline{{\it New York University, New York, NY, 10003, USA}}
\end{center}

\begin{abstract}
As is well known, both massive gravity and bigravity exhibit the linear van Dam-Veltman-Zakharov (vDVZ) discontinuity  
that is cured classically by the nonlinear Vainshtein mechanism due to certain low scale strongly coupled 
interactions. Here we show how both the vDVZ and strong coupling problems 
can be removed by embedding 4D covariant massive gravity into a certain 
5D warped geometry. The 4D theory is a nonlinear strongly coupled massive gravity, that is being coupled to 
a 5D bulk theory that generates a bulk graviton mass via a one loop diagram.  This induced mass leads to 
an additional 4D kinetic term for the 4D longitudinal mode, even on flat space. Due to this kinetic term the 4D  massive 
theory becomes weakly coupled all the way up to a high energy scale set by the bulk 
cosmological constant.  The same effect leads to a suppression of the  interactions of the 4D longitudinal  
mode with a 4D matter stress-tensor, thus removing the vDVZ discontinuity. 

\vskip 0.15cm
\noindent
The proposed mechanism has a pure 4D holographic interpretation: a 4D nonlinear massive gravity mixes  
to a non-conserved symmetric tensor of a 4D CFT that has a cutoff; the latter mixing generates a large kinetic term 
for the longitudinal mode, and  this makes the longitudinal mode be weakly coupled to a matter stress-tensor, 
and weakly self-coupled, all the way up to the scale of the 4D CFT cutoff.

\end{abstract}

\newpage

\tableofcontents

\section{Introduction and toy models}
\label{intro}
The modern view of General Relativity (GR) is that it is the unique interacting theory of a massless graviton, valid at distances larger than its short distance cutoff -- the Planck length.~While it is certainly to be completed below this length scale, GR is in excellent agreement with observations at longer distances, ranging from sub-milimeter scales, and all the way up to the cosmological ones.~Nevertheless, inability to provide a compelling quantum field theory explanation to the smallness of the cosmological constant, as well as (perhaps relatedly) to the origin of the late-time  cosmic acceleration has motivated some to entertain the possibility that GR is modified  also at long distances, of order of the size of the observable Universe. 

\vskip 0.15cm
\noindent
Perhaps the simplest  modification of GR corresponds to positing that the graviton has a nonzero mass; the latter can be chosen, in a technically natural way, to be of order $m \sim 10^{-33}\,\rm{eV}$, the Hubble scale today.~Naively, this would correspond to modifying  the theory around the graviton's Compton wavelength (of order of the current size of the Universe), at the same time leaving gravity and all of its empirical success unaffected at shorter distances.~This conclusion is too quick, however.~A massive graviton, no matter how light, has three more degrees of freedom compared to the massless one, and one of those three---the graviton's scalar longitudinal polarization $\pi$---tends to introduce various peculiarities in the dynamics of the theory \cite{vanDam:1970vg, Zakharov:1970cc, Vainshtein:1972sx, Dvali:2000hr, Deffayet:2001uk, Luty:2003vm, ArkaniHamed:2002sp}.~Most importantly, in special, ghost-free theories of massive gravity \cite{deRham:2010ik, deRham:2010kj}, the interactions of $\pi$ become strongly coupled at a rather low energy scale
\be
\Lambda_3 = (M_{\rm Pl} m^2)^{1/3}\sim (10^3\,\rm{km})^{-1}\,.
\ee
The strong coupling scale is even lower in a \textit{generic} local and Poincar\'e-invariant theory of massive gravity \cite{ArkaniHamed:2002sp}.

\vskip 0.15cm
\noindent 
Strong coupling of the graviton's scalar polarization is both a blessing and a curse.~On the one hand, it leads to the Vainshtein mechanism, the non-linear screening of $\pi$'s contribution to the \textit{classical} Netwonian potential, and guarantees that the predictions of massive gravity are in agreement with those of GR below a certain macroscopic length scale, known as the Vainshtein radius \cite{Vainshtein:1972sx, Deffayet:2001uk}.~On the other hand, once the theory is treated \textit{quantum mechanically}, the strongly-coupled dynamics of $\pi$ lead to the loss of perturbative unitarity and validity of the classical approximation below distances of order $\Lambda_3^{-1}$ becomes dependent on assumptions about the unknown ultraviolet (UV) comptetion of the theory \cite{Nicolis:2004qq}. 

\vskip 0.15cm
\noindent
This phenomenon of strong coupling is by no means unfamiliar.~It is in fact fully analogous to what occurs in low-energy theories of (non-Abelian) massive spin-1 particles, e.g. $W$ bosons.~In addition to the two transverse polarizations of its massless counterpart, a $W$ boson of mass $m_W$ and gauge coupling $g$ features an additional, longitudinal mode which becomes strongly coupled at a scale $4\pi m_W/g$ -- the spin-1 analog of $\Lambda_3$.~Scattering longitudinal $W$ bosons above this energy scale can only be studied after specifying the short-distance completion of the theory.~In the Standard Model (SM) of particle physics, such a (weakly coupled) short-distance completion is of course provided by the Higgs particle, which comes in at the scale $m_H \ll 4\pi m_W/g$ and `unitarizes' the scattering amplitudes by cancelling the pieces that grow with the center of mass energy.
\vskip 0.15cm
\noindent
Can the low-energy theory of a massive graviton be UV completed along the lines of the traditional Higgs mechanism of the SM? Interestingly, for the theory formulated on flat spacetime, the answer appears to be negative.~Among other evidence, it has been recently shown that it is \textit{impossible} to improve the high-energy behaviour of the tree-level longitudinal graviton scattering by including the exchange of any number of vector and scalar particles \cite{Bonifacio:2019mgk}.\footnote{With higher spins, this conclusion does not apply, of course:~the obvious exceptions are provided by string theory, or higher-dimensional GR with compactified spatial dimensions.~The (explicitly unitary) $4D$ effective descriptions of these theories invove massive spin-2 particles in both cases.~However, in these theories there is no parametric separation between the mass of the lightest such particle and the masses of heavier states (string theory resonances or KK modes respectively).~We will be exclusively interested in theories in which the lightest massive spin-2 particle can be parametrically isolated on the energy scale.}
 
\vskip 0.15cm
\noindent
Incompatibility of flat-space massive gravity with a traditional, weakly coupled Higgs mechanism can already be grasped in the effective theory below the $\Lambda_3$ scale.~To that end, note that (in a certain window of energies below $\Lambda_3$) the full massive spin-2 multiplet can be written as the following combination
\be
h_{\mu\nu} - \p_{(\mu} V_{\nu)}\,,
\label{helicitysplit}
\ee
where $h_{\mu\nu}$ describes the 2 degrees of freedom of the transverse general-relativistic (helicity-2) graviton, while $V_{\mu}$ encode the three extra polarizations that the graviton needs to `eat up' to become massive.~These degrees of freedom presumably come from some `Higgs' sector of the theory, analogous to the complex Higgs doublet of the SM.~At high energies---those well above $\Lambda_3$---the standard picture would consist of this Higgs sector, coupled to general relativity (corresponding to the Coulomb phase of the theory with a massless graviton).~At lower energies, the Higgs sector  
provides the three Nambu-Goldstone (NG) bosons to be "eaten up" by the graviton (possibly along with extra `radial modes', responsible for unitarizing the longitudinal graviton scattering around the $\Lambda_3$ scale \textit{a la} the standard model Higgs).

\vskip 0.15cm
\noindent
Alas, such a simplistic picture for a UV theory of massive  gravity does not seem plausible.~This can be deduced by plugging the decomposition \eqref{helicitysplit} into the linearized action, and zooming onto distances, much smaller than the graviton's Compton wavelength -- a regime known as the decoupling limit of the theory.~In the given regime, the action should split into two separate sectors, one describing massless general relativity (represented by $ h_{\mu\nu}$), and the other -- the three NG bosons in the Higgs sector.~However, one immediately observes that the NG action is degenerate in this limit:
\be
S_V  =  \mpl^2 m^2 \int d^4 x\, \( -\frac{1}{4} F^{\mu\nu} F_{\mu\nu} \) \,,\qquad F_{\mu\nu} = \p_\mu V_\nu - \p_\nu V_\mu\,,
\ee
propagating \textit{two} degrees of freedom, instead of three.~The missing degree of freedom -- the longitudinal scalar ($V_\mu\sim \p_\mu\pi$) only acquires dynamics upon reintroducing coupling to gravity, via a kinetic mixing of the form $\mathcal{L}\sim h\p^2\pi$.~One is thus led to conclude that the putative Higgs sector that would complete massive gravity at short distances  is itself ill-defined in the limit of decoupled gravity.~Or,  to put it differently, the theory does not seem to admit a well-defined Coulomb phase, characterized by the presence of a \textit{massless} graviton.\footnote{None of these conclusions apply to massive non-Abelian spin-1 theories, of course.~One can easily check that in the analogous limit (known as the Goldstone equivalence limit \cite{Chanowitz:1985hj}), the low-energy action for such particles splits into separate, \textit{non-degenerate} sectors, describing the transverse and longitudinal/NG polarisations of the vector boson.}

\vskip 0.15cm
\noindent
It is precisely the absence of an independent kinetic term for $\pi$ that leads to the low strong coupling scale of massive gravity, as well as the van Dam-Veltman-Zakharov (vDVZ) discontinuity of the linear theory \cite{vanDam:1970vg, Zakharov:1970cc}.  
One could try to modify the action of the theory in a way that gives rise to such a kinetic term; however, it is possible to show that this cannot happen within a unitary (ghost-free) and local field theory on flat space.

\vskip 0.15cm
\noindent
The degeneracy of the NG sector can be eliminated on \textit{curved} space, however.~Indeed, it is known that a conventional Higgs mechanism exists for anti de Sitter (AdS) gravity \cite{Porrati:2000cp, Porrati:2001db}.~From the low-energy perspective, this can be seen by noting that unlike its flat-space counterpart, the NG vector $V_a$ of a massive AdS graviton acquires non-zero mass, stemming from non-zero curvature of the background spacetime:
\be
\label{VonAds}
S_V = M^{d-1}_{d+1} m^2 \int d^{d+1}x\, \sqrt{\bar g} \(-\frac{1}{4}F^{ab} F_{ab} - \frac{d}{L^{2}}V^a V_a\)\,.
\ee
Here $\bar g_{ab}$ denotes the  background metric of AdS (used to raise/lower all indices), $d+1$ and $L$ are respectively the spacetime dimensionality and curvature radius, and $M_{d+1}$ is the $(d+1)$-dimensional Planck mass.~(Notice that the mass of the NG vector is always $m^2_V = 2 dL^{-2}$, regardless of the precise value of the graviton's mass.)~With the mass term present in the action \eqref{VonAds}, the longitudinal scalar $\pi$ does now acquire an independent kinetic term, and the massless (decoupling) limit of the theory is continuous, describing two separate sectors, one consisting of massless gravity and the other -- of the three dynamical NG modes. 

\vskip 0.15cm
\noindent
The above linearized-level mechanism for generating a kinetic term for $\pi$ on AdS can be embedded into a nonlinear theory, realizing a full-fledged  gravitational Higgs mechanism \cite{Porrati:2001db, Porrati:2003sa}.~The full (UV) theory is of a rather conventional type: standard general reativity, coupled to scalar fields with specific boundary conditions.~The ultraviolet cutoff of this theory is at least of order of the AdS curvature, and can be very large, e.g. around the GUT scale: $L^{-1}\sim 10^{16}\,\rm{GeV}$ (we will assume this is somewhat lower than the $d+1$ dimensional Planck mass $M_{d+1}$, to have a weakly-coupled description of AdS gravity).~We will see that the graviton mass in this theory comes out to be suppressed by the Planck scale, as well as an extra dimensionless coupling constant $\lambda$ (which can be naturally arbitrarily small):
\be
\bar m^2\sim \frac{\lambda^2}{M^{d-1}_{d+1} L^{d+1}} \ll L^{-2}\,.
\ee
 \vskip 0.15cm
\noindent
However, we cannot directly use the nice properties of AdS massive gravity for phenomenological purposes, as the spacetime we live on is better approximated by Minkowski space, rather than anti de Sitter.~Instead, we will imagine that our 4D flat Universe forms the boundary of a 5-dimensional AdS bulk, parametrized by the coordinates $x^\mu$ (our 4D spacetime) and $z$ (the 5th dimension).~Moreover, we will assume that the gravitational action describes non-linear, ghost-free dRGT gravity, confined to the 4D boundary at the origin of the $z$-axis, as well as a \textit{massive} 5D bulk graviton, whose mass arises from the above-described Higgs mechanism in GR, coupled to scalars on anti de Sitter space.~Without the bulk, the boundary theory would be strongly coupled at the $\Lambda_3$ scale, as discussed above; however, we'd like to argue that the bulk dynamics drastically changes the state of affairs, raising the strong coupling scale by many orders of magnitude. 

One should perhaps note at this point that at the non-linear level, the resulting theory of 5D massive gravity \textit{does not} belong to the dRGT class.~The light graviton is accompanied by a tower of `bound states' (made out of the two `fundamental' scalars we started with) of various spins with AdS `masses' of order $L^{-1}$ or larger.  Each of these states induces a continuum tower of gapless 4D modes; hence it might be more useful to think of the 5D massive graviton as possessing a certain form factor characterized by the curvature scale.~Nevertheless, at distances much larger than $L$, the relevant part of the theory reduces to the standard Fierz-Pauli (free) massive gravity, and this will be the only part of the action that we will need to use for our purposes.

\vskip 0.15cm
\noindent
Coming from a well-defined Goldstone sector, the scalar longitudinal mode of the \textit{bulk} graviton, denoted here by $\Pi(x,z)$, does have an independent $5D$ kinetic term in the proper short-distance limit of the theory.~This, as we will show below, allows also the 4D helicity-0 graviton, related to $\Pi$ simply as $\pi(x) =\Pi(x,0)$, to acquire a quadratic action that resembles a certain non-local kinetic term.~This 4D kinetic term can be thought of as arising from `integrating out' a gapless continuum of states (Kaluza-Klein modes of higher-dimensional gravity) and it is precisely its non-locality that allows to evade the difficulties associated with giving independent dynamics to $\pi$ within a unitary and Poincar\'e invariant theory of massive gravity on flat space \cite{Gabadadze:2017jom}.

\vskip 0.15cm
\noindent
Nevertheless, at momenta much lower than the AdS curvature, the bulk-induced 4D effective action for $\pi$ can be approximated (in a rather subtle way, discussed below) by an ordinary, local kinetic term. This makes the 4D boundary dynamics of this field much more weakly coupled than it would be in the absence of the bulk.~In fact, we will find that the resulting $4D$ theory can be weakly coupled all the way up to energies of order $L^{-1}$.~For a GUT-scale bulk curvature, this corresponds to raising the strong coupling scale of massive gravity by some 38 orders of magnitude.

\vskip 0.15cm
\noindent
The mechanism by which non-local interactions can alleviate strong coupling of longitudinally polarized bosons can be grasped by studying a toy model, describing a massive spin-1 particle $A_\mu$ on flat space:
\be
\mathcal{L} = -\frac{1}{4} F^2_{\mu\nu} -\frac{1}{2} m^2 A^2_\mu - \frac{\lambda}{4} (A^\mu A_\mu)^2 + \mathcal{L}_{\rm mix}\,.
\ee
The last term describes a certain  (non-local) mixing/interaction with an external sector, to be specified in a moment.~Without this term, the canonically-normalized longitudinal mode enters the spin-1 multiplet as $A_\mu \sim m^{-1} \p_\mu \hat\pi$, and is strongly coupled at a scale $\Lambda_{\rm loc} \sim \lambda^{-1/4}\, m \,.$\footnote{For small $\lambda$, this scale is larger than $m$, so the ultra-relativistic regime can still be captured by the low-energy effective theory of $A_\mu$.} 

\vskip 0.15cm
\noindent
We'd  like the mixing, described by $\mathcal{L}_{\rm mix}$ to generate an additional right-sign kinetic term for the longitudinal mode, in order to make it more weakly coupled.~A  kinetic  mixing 
of $\hat\pi$ with another scalar would generate an additional kinetic term for the former field; however, it would have a wrong sign after diagonalization.~One way to generate a right-sign kinetic term is to mix the longitudinal  mode with a  scalar that itself 
has a wrong-sign kinetic term (i.e.~a scalar ghost).~To avoid ghost instabilities, however, the latter scalar has to be non-propagating, which is the case if it is a conformal mode of a massless tensor field, constrained by the structure of general relativity.~In light of these 
arguments we will consider the following interaction/mixing with an external massless spin-2 field $\chi_{\mu\nu}$\,:
\be
\label{chiAmixing}
\mathcal{L}_{\rm mix} = -\frac{1}{4}\, \chi^{\mu\nu} (\hat{ \mathcal{E}} \chi)_{\mu\nu} - \frac{M}{2}\, \chi^{\mu\nu}\, \(\frac{\p_\mu\p_\nu}{\p^2} - \eta_{\mu\nu}\)\p^\alpha A_\alpha\,.
\ee
Here $\hat{ \mathcal{E}}$ in the kinetic term of $\chi$ denotes the qudratic Einstein-Hilbert  operator, and the whole action is invariant under gauge transformations of the form $\chi_{\mu\nu} \to \chi_{\mu\nu} +  \p_{(\mu}\xi_{\nu)}$, which  guarantees that the spin-2 boson propagates no more than two (transverse-traceless) degrees of freedom.~Notice that the projector in the last term extracts $\chi$'s conformal mode.~Integrating out the massless $\chi$ results in a non-local contribution to the vector action:
\be
\label{Aeft}
\mathcal{L}_A = -\frac{1}{4} F^2_{\mu\nu} -\frac{1}{2} m^2 A^2_\mu - \frac{\lambda}{4} (A^\mu A_\mu)^2 + \frac{3}{2}\, M^2\, \p^\mu A_\mu \, \p^{-2} \,\p^\nu  A_\nu\,,
\ee
and the important point is that the new non-local piece gives rise to a (right-sign!) kinetic term for the vector's longitudinal mode $\pi$.~For a non-canonically normalized $\pi$ defined as $A_\mu \sim \p_\mu\pi$, the full action reads
\be
\mathcal{L}_\pi = -\frac{1}{2} \(m^2 + 3 M^2\)(\p\pi)^2 - \frac{\lambda}{4} (\p\pi)^4\,,
\ee
and as far as $M\gg m$, the propagation of $\pi$ is fully determined by the non-local piece in \eqref{Aeft}.~The strong coupling scale in this case is $\Lambda_{\rm nloc} = \lambda^{-1/4}\, M \gg \Lambda_{\rm loc}$. Furthermore, if we were to introduce a coupling of the longitudinal mode to  some other source, $\mathcal{L}_A\sim \p^\mu A_\mu\, J$, this  coupling would also be further suppressed due to the rescaling of the $\pi$ kinetic term.~Thus, the theory has become more weakly coupled due to the (non-local) mixing with the `hidden' sector containing a tensor $\chi$- field. 

\vskip 0.15cm
\noindent
A similar phenomenon of kinetic term enhancement for the longitudinal mode of a massive Fierz-Pauli tensor $h_{\mu\nu}$ results from its mixing with the conformal mode of the massless spin-2 $\chi$-field, via the following term in the Lagrangian (up to a coefficient)
\be
\chi^{\mu\nu}\, \(\frac{\p_\mu\p_\nu}{\p^2} - \eta_{\mu\nu}\)\p^{-2} \(\p^\alpha \p^\beta h_{\alpha\beta} - {1\over 4} \p^2 h\).
\label{chihmixing}
\ee
Needless to say, a \textit{generic} kind of non-locality would make it hard to reconcile the theory with basic QFT principles, such as causality.~However, the four-dimensional effective non-locality we will rely on is very special: it stems from a fully local higher-dimensional AdS theory, in which no issues with causality arise by the very definition.

\vskip 0.15cm
\noindent
Going back to the proposed theory of massive gravity, we stress that there are two fully equivalent ways of thinking about the above-described mechanism of scaling up the strong coupling.~The local 5D gravity picture seems more tractable, but it can also be understood, via the AdS/CFT correspondence, in purely four-dimensional terms.~In the 4D formulation, the strong coupling (energy) scale is raised as a result of coupling the massive graviton to a conformal sector, dual to the continuous Kaluza-Klein spectrum of extra-dimensional gravity \cite{Gabadadze:2017jom, Gabadadze:2006jm}.

\vskip 0.15cm
\noindent
The rest of the paper is organized as follows.~Sec.~\ref{summary} provides a summary of the proposed construction, with a quick description of the basic mechanism that allows to significantly raise the strong coupling scale of massive gravity, confined to a flat brane in AdS$_5$.~The schematic discussion of Sec.~\ref{summary} is then expanded upon in Sec.~\ref{sec3}, where we provide a much more detailed account of the theory and its dynamics.~Finally, in Sec.~\ref{cft} we discuss the dual CFT interpretation of our gravitational setup.~Some of the technical details that would overload the main presentation are relegated to the two appendices.

\section{A summary of the proposal and results}
\label{summary}
The purpose of this section is to provide a summary of the mechanism by means of which the UV cutoff of 4D massive gravity is raised from $\Lambda_3 \sim 10^{-22}~\rm{GeV}$ to a new scale $L^{-1}$, which can be as high as $L^{-1} \sim 10^{16}~\rm{GeV}$.~A longer and a more detailed discussion of the setup is the subject of the next section, which together with the present one contains the main results of the paper.

\vskip 0.15cm
\noindent
We will assume that our world is confined to a brane, embedded in a slice of 5D anti de Sitter space of curvature $L^{-2}$, described by the following metric
\be
\label{regads}
ds^2 \equiv \bar g_{ab}dx^a dx^b = \frac{L^2}{(|z|+L)^2}\(\eta_{\mu\nu} dx^\mu dx^\nu + dz^2\).
\ee
The fifth dimension ranges from $z=0$ (where our brane is located)  to $z=\infty$, corresponding to the AdS horizon.~Equivalently, one can think in terms of a two-sided setup, with the brane located at an orbifold fixed point which separates \textit{two} slices of anti de Sitter space, related to each other by a $\mathbb{Z}_2$ symmetry under $z\to -z$ (hence the absolute value in eq.~\eqref{regads}).\footnote{In this case one can think of $z=0$ as the position of a physical brane (e.g. arising from some kind of a solitonic configuration at the microscopic level), embedded in an unbounded spacetime.~It is then clear that one does not have to worry about adding boundary terms such as the Gibbons-Hawking term or its massive gravity generalization \cite{Gabadadze:2018bpf} to the  brane action.}~(Technically, the two setups are fully equivalent, apart from factors of 2 that show up in some of the expressions.)~The metric \eqref{regads} describes a regularized version of anti de Sitter, which in the standard Poincar\' e coordinates defined as
\be
\label{zprime}
z' = L + z \,,
\ee
corresponds to cutting the spacetime off a small distance $\epsilon$ away from the (would-be) boundary $z' = 0$ of the complete Poincar\' e patch; sending $\epsilon\to 0$ corresponds to the AdS/CFT limit.~In our setup, $\epsilon = L$, which is indeed small compared to other scales we will be interested in.

\subsection{The gravitational action}
The complete gravitational action of the proposed theory consists of the four-dimensional (brane) and the five-dimensional (bulk) parts
\be
\label{brane+bulk}
S = S_{brane } + S_{ bulk}\,,
\ee
where the two contributions are given respectively in eqs.~\eqref{sbdy} and \eqref{bulkaction1} of the next section.~Both describe massive gravity (respectively in four and five dimensions), although the corresponding theories differ in  nature.~The theory on the brane is dRGT massive gravity.~The bulk action, on the other hand, describes conventional general relativity, coupled to a pair of conformal scalars with specific boundary conditions that give rise to a gravitational Higgs mechanism on the AdS background of the theory.~Both the brane and the bulk theories are discussed in great detail in the next section (respectively in sections~\ref{drgt} and~\ref{higgsonads}-\ref{bigravity}).

\vskip 0.15cm
\noindent
In its covariant formulation, the brane (dRGT) theory reduces to  general relativity, coupled in a specific way to four diffeomorphism scalars $\phi^p$ ($p = 0\dots 3$).~In a certain high-energy regime, the fluctuations of these scalars about their expectation values, $\langle \phi^p \rangle = \delta^p_\mu x^\mu$, describe the extra (helicity-1 and helicity-0) polarizations that a 4D massive graviton propagates in addition to its two general-relativstic (helicity-2) polarizations.

\vskip 0.15cm
\noindent
In the parametrically large window of energies $m \ll E \ll \Lambda_3$ defining what's known as the \textit{decoupling limit} of the theory, the most interesting dynamics of massive gravity are captured by an action of the following (schematic) form \cite{deRham:2010ik}
\be
\label{schematic}
\mathcal{S}^{dl}_{brane}\sim \int_{brane} \mpl^2\, h\p^2 h -\mpl^2 m^2 \big[h\p^2\pi + \beta_2 h (\p^2\pi)^2 + \beta_3 h (\p^2\pi)^3 \big] + h^{\mu\nu}T_{\mu\nu}\,,
\ee 
where $\pi$ denotes the longitudinal scalar polarization of the massive graviton, $h_{\mu\nu}$ describes its helicity-2 part and $T_{\mu\nu}$ is the stress tensor of matter.~The first term in \eqref{schematic} is the linearized Einstein-Hilbert action, while the next three (those in the parentheses) originate from the mass/potential terms of the graviton.~The most general unitary theory of 4D massive gravity is characterized, in addition to the graviton's mass, by two constant parameters, $\beta_2$ and $\beta_3$.~Apart from the last term, we have neglected the index structure in \eqref{schematic} (see a more detailed discussion around eq.~\eqref{dlactionfinal} below), but we note that it is of a very special type, leading to dynamical equations of at most second order in time, despite the presence of higher-derivative interactions.~Moreover, in the given (non-trivial, interacting) limit, the theory is \textit{exactly} invariant under linearized diffeomorphisms
\be
\label{lineardiffs}
\delta h_{\mu\nu} = \p_\mu \xi_\nu + \p_\nu\xi_\mu\,,
\ee
which is due to the fact that the `currents' made out of $\p^2\pi$ that $h_{\mu\nu}$ couples to are kinematically conserved.

\vskip 0.15cm
\noindent
As discussed in Sec.~\ref{intro}, the longitudinal mode $\pi$ has no dynamics `on its own'; however, once its quadratic mixing with the helicity-2 polarization $h_{\mu\nu}$ is taken into account (the second term in~\eqref{schematic}), $\pi$ acquires a kinetic term of the form
\be
\label{4dpikinterm}
 \mpl^2 m^4 \pi\p^2\pi\,.
\ee
This term arises as a result of switching to a new field basis $h_{\mu\nu}\to h_{\mu\nu} + \pi \eta_{\mu\nu}$, in terms of which the quadratic part of the action is diagonalized.~At the same time, the latter field redefinition generates a coupling of $\pi$ to the trace of the matter stress tensor:
\be
\label{lpit}
\mathcal{L}_{\pi T} =  m^2 \pi T.
\ee
Canonically normalizing the fields, $h_{\mu\nu}\to \mpl^{-1}h_{\mu\nu}$ and $\pi\to (\mpl m^2)^{-1} \pi$, and carefully inspecting the action reveals that the interactions of $\pi$ become strong at the low scale $\Lambda_3$, while the coupling of $\pi$ to matter is order-one in units of $\mpl^{-1}$, leading to a `fifth force' of gravitational strength.~To avoid conflict with observations, one needs to rely on precisely the strongly-coupled dynamics of $\pi$, which suppresses the contribution of this field to the classical Newtonian potential \cite{Vainshtein:1972sx, Deffayet:2001uk}.~Nevertheless, $\Lambda_3$ is the true quantum cutoff of the theory and predictivity of massive gravity at characteristic distance scales below $\Lambda_3^{-1}\sim 10^3\, \rm{km}$ relies on making certain assumptions about the putative short-distance completion of the dynamics \cite{Nicolis:2004qq}.

\vskip 0.15cm
\noindent
We wish to argue that these issues are remedied upon inclusion of the bulk part of the action~\cite{Gabadadze:2017jom}.~While the bulk theory is described in detail in the next section, we note that at energies and momenta lower than $L^{-1}$, it effectively reduces to the 5D Fierz-Pauli theory of a free AdS graviton $\bar h_{ab}$ with mass $\bar m \ll L^{-1}$.~The (linearly) diffeomorphism-invariant action of this theory, including the bulk \stu vector $V_a$\,, reads 
\begin{align} \label{FP1}
S_{ bulk} &\simeq M^3_{5} \int d^{5} x \, \sqrt{-\bar g}\, \Big[ -\frac{1}{4} \nabla_{c} \bar  h_{ab} \nabla^{c} \bar  h^{ab} +\frac{1}{2} \nabla_{c} \bar  h_{ab} \nabla^{b} \bar h^{ac} -\frac{1}{2} \nabla_{a}\bar  h \nabla_{b}\bar  h^{ab} + \frac{1}{4} \nabla_{a} \bar  h \nabla^{a} \bar  h \nonumber \\
&- \frac{2}{ L^2} \Big( \bar h_{ab}\bar  h^{ab} - \frac{1}{2} \bar  h^2 \Big) -\frac{\bar m^2}{4}\(\(\bar h_{ab}- 2\nabla_{(a}V_{b)}\) ^2- (\bar  h - 2\nabla\cdot V)^2\)\Big]\,,
\end{align}
where $\nabla$ denotes the covariant derivative with respect to the background metric $\bar g_{ab}$.~Carefully accounting for the non-trivial commutation rules of AdS covariant derivatives reveals that $V_a$ is a \textit{massive} vector with mass $m_V^2 = 8 L^{-2}$, in agreement with eq.~\eqref{VonAds} of the introductory section.

\vskip 0.15cm
\noindent
From now on, we will concentrate on the helicity-2 ($\bar h_{ab}$) and the helicity-0 ($V_a \sim \nabla_a\Pi$) components of the bulk graviton, and zoom onto distance scales much smaller than the graviton's compton wavelength, $\bar m \to 0$, which defines the decoupling limit of the bulk theory:
\be
\label{bulkdl}
S^{dl}_{ bulk } \sim \int_{ bulk} M^3_5  \nabla \bar h \nabla \bar h - M^3_5 \bar m^2 L^{-2} (\nabla\Pi)^2
\ee
The first term in this expression schematically denotes the 5D Einstein-Hilbert term, expanded to the quadratic order on anti de Sitter space, while the second is the kinetic term for $\Pi$, which arises thanks to the vector $V_a$ being massive on AdS.~As discussed above, the latter kinetic term would be absent on flat space, and the dynamics of $\Pi$ would only arise through mixing with $\bar h$; in the present case of AdS background, however, this mixing is completely negligible.~The theory \eqref{bulkdl} is exactly invariant under the linearized bulk diffeomorphisms
\be
\label{bulkdiffs}
\delta \bar h_{ab} = \nabla_a \bar \xi_b +  \nabla_b \bar \xi_a\,,
\ee
however, as we will be exclusively working in the gauge corresponding to an unbent brane at $z=0$, we have to require $\bar \xi_z (x, 0) = 0 $.~It is in this gauge that we identify the boundary fields entering the brane theory~\eqref{schematic} with their bulk counterparts:
\be
\label{fieldid}
h_{\mu\nu} (x) = \delta_\mu^a \delta_\nu^b\, \bar h_{ab}(x, 0)\equiv \bar h_{\mu\nu}|  \,, \qquad \pi(x) = \Pi|\,.
\ee
From now on, a vertical stroke will denote evaluation on the brane.~Keeping the brane at $z=0$, one can further fix the gauge so that $\bar h_{zz} = \bar h_{\mu z} = 0$ is true in the bulk.~The residual freedom then corresponds to choosing the bulk gauge parameters as $\bar \xi_\mu (x, z)= L^2 (z + L)^{-2} \,\omega_\mu(x)\,,~\bar\xi_z (x, z) = 0$, which generates the four-dimensional  brane diffeomorphisms \eqref{lineardiffs}, corresponding to $\xi_\mu(x) = \omega_\mu(x)$ \cite{Gabadadze:2017jom}.

\subsection{Dynamics}
Let us summarize our setup.~The simplified model that we wish to explore in the rest of this section is specified by the total (bulk + brane) decoupling limit action 
\begin{align}\label{fulldl}
S^{dl} &=  \int_{brane} \mpl^2\, h\p^2 h -\mpl^2 m^2 \big[h\p^2\pi + \beta_2 h (\p^2\pi)^2 + \beta_3 h (\p^2\pi)^3 \big] + h^{\mu\nu}T_{\mu\nu} \nn \\
&+  \int_{ bulk} M^3_5  \nabla \bar h \nabla \bar h - M^3_5 \bar m^2 L^{-2} (\nabla\Pi)^2\,,
\end{align}
supplemented by the identification \eqref{fieldid} of the brane and the bulk fields.~The bulk theory is understood to be gauge-fixed so that the brane is unbent at $z=0$ (that is, the `brane bending mode' is gauged away).~The action is then invariant under the linearized bulk diffeomorphisms \eqref{bulkdiffs} with $\bar\xi_z |= 0$, and the corresponding 4D reparametrizations of the brane \eqref{lineardiffs}. One can use this freedom to impose $\bar h_{zz} = \bar h_{\mu z} =0$ in the bulk, which still leaves residual gauge invariance, under which the bulk and the brane fields transform with \{$\bar \xi_\mu (x, z) = L^2 (z+L)^{-2} \omega_\mu(x),~\bar\xi_z = 0$\} and $\xi_\mu(x) = \omega_\mu(x)$.

\vskip 0.15cm
\noindent
The dynamical equations for $\bar h_{ ab}$ and $\Pi$ that follow from varying the bulk action \eqref{fulldl} describe respectively a massless spin-2 field (fully analogous to the graviton of general relativity) and a massless scalar on anti de Sitter space.~It is well-known that the massless tensor field gets \textit{localized} on a positive tension brane in AdS \cite{Randall:1999vf}, and so does a massless scalar \cite{Bajc:1999mh}.~That is, even if one does not include the `bare'  action $S_{brane}$ to start with, their brane images $h_{\mu\nu}(x)$ and $\pi(x)$ acquire 4D kinetic terms, thereby mediating four-dimensional interactions between brane sources.

\vskip 0.15cm
\noindent
With the given normalization of modes, the induced 4D kinetic terms are of the following form \cite{Randall:1999vf, Bajc:1999mh}\footnote{There is also an induced kinetic mixing between $h_{\mu\nu}$ and $\pi$ \cite{Gabadadze:2017jom}, but this won't be important in the following discussion.}
\be 
\label{linduced}
S^{dl}_{ind} \sim \int_{brane} M^3_5 L h\p^2h +\frac{M^3_5\bar m^2}{L}\pi\p^2\pi \,,
\ee 
and the complete 4D effective action is obtained by adding $S^{dl}_{ind}$ to the bare brane action $S^{dl}_{brane}$\,.~To further simplify the discussion, we will assume that the bulk and the brane masses are of the same order, $m \sim \bar m$.~Moreover, the bare and the induced Planck masses will also be assumed to be set by the same scale: $M^3_5 L \sim \mpl^2$.~One then finds that the first term in the brane-induced action \eqref{linduced} leads to an additive renormalization of the coefficient of the already existing (Einstein-Hilbert) term in the bare brane world-volume action \eqref{schematic}.~In contrast, the $\pi$ kinetic term, induced from the bulk is a genuinely novel feature of the effective theory on the brane, which now has the following form
\be
\label{braneplusbulk}
S^{dl} \sim\int_{brane}  \mpl^2 h\p^2 h + \frac{\mpl^2 m^2}{L^2}\pi\p^2\pi - \mpl^2 m^2 \big[h\p^2\pi + \beta_2 h (\p^2\pi)^2 + \beta_3 h (\p^2\pi)^3 \big] + h^{\mu\nu}T_{\mu\nu}\,.
\ee
Given that $m^2 \ll L^{-2}$, the induced kinetic term of $\pi$ is much larger than the kinetic term \eqref{4dpikinterm} that results from mixing with the helicity-2 mode.~With the bulk physics taken into account therefore, this mixing is irrelevant for $\pi$ propagation, and the canonically normalized fields become $\hat h_{\mu\nu} = \mpl\, h_{\mu\nu}$, and $\hat\pi = \mpl L^{-1} m\,\pi \,.$

\vskip 0.15cm
\noindent
The small mixing with the helicity-2 polarization does still give rise to a coupling of $\hat \pi$ to the matter stress tensor of the form
\be
\label{piTcouplig}
\mathcal{L}_{\pi T} =\frac{\alpha}{\mpl}\,\hat \pi T \,,
\ee
with $\alpha \sim mL$.~Unlike purely 4D massive gravity, however, this coupling is tiny as the graviton's mass is parametrically smaller than the AdS curvature scale.~This renders the fifth forces, mediated by the helicity-0 graviton very small, and in fact, there is no vDVZ discontinuity \cite{vanDam:1970vg, Zakharov:1970cc} even on flat space -- the graviton mass can be taken to zero and  with it,  the coupling of the longitudinal mode to a matter stress-tensor goes to zero as well. 

\vskip 0.15cm
\noindent
The enhanced kinetic term of the helicity-0 graviton results in the weak coupling of the theory all the way down to distances well below any macroscopic scale.~To show this, one can examine the most relevant interactions of $\hat \pi$ which, as it turns out, are given by the following terms
\be
\label{lambdadefs}
S^{dl} \sim\int_{brane} \hat \pi\p^2 \hat \pi - \frac{\beta_2}{M_\star^3 }\, \epsilon\epsilon \hat h  (\p^2\hat\pi)^2 - \frac{\beta_3}{\Lambda_\star^6 }\, \epsilon\epsilon \hat h  (\p^2\hat \pi)^3 + \dots\,,
\ee
where the suppression scales have been defined as 
\begin{align}
\label{lambdas}
M^3_\star \sim \frac{\mpl}{L^{2}} \gg \frac{1}{L^3}\,, \qquad
\Lambda^6_\star  \sim \frac{\mpl^2 m}{L^3}\,.
\end{align}
The second of these scales, $\Lambda_\star$, is lower than $M_\star$, although it is still much higher than $\Lambda_3$\,. Indeed, for a GUT-scale AdS curvature and a Hubble-scale graviton mass, we have $\Lambda_\star \sim 10^7\, \rm{GeV}$, to be compared with $\Lambda_3 \sim 10^{-22}\, \rm{GeV}$: the cutoff of the brane theory has increased by 29 orders of magnitude.

\vskip 0.15cm
\noindent
Furthermore, setting $\beta_3 = 0$---a technically natural choice in the theory at hand \cite{deRham:2012ew}---results in further increase in the strong coupling (energy) scale.~In this case, the scale suppressing the strongest interaction is $M_\star$.~This scale is greater than AdS curvature, however, and it turns out not to have physical meaning -- the true cutoff of the 4D theory is set by $L^{-1}$.~This is because, as we discuss in more detail in the next section, the latter scale marks the point where new \textit{bulk} states---the `radial Higgs' modes involved in the 5D gravitational Higgs mechanism---start to become `visible' to the 4D brane observer. 

\vskip 0.15cm
\noindent
For the most general choice of the parameters  at hand (that is, without assuming $m\sim \bar m$ and $M^3_5 L\sim \mpl^2$), the expressions for the two scales $M_\star$ and $\Lambda_\star$ generalize to those given in eq.~\eqref{lambdasgeneral} of the next section.

\vskip 0.15cm
\noindent
We have stated above that it is impossible to modify the Nambu-Goldstone sector of massive gravity within a unitary and local theory, and yet we have argued that a consistent modification exists which, at least in some limit, gives rise to a local kinetic term for $\pi$.~How is that possible?~The resolution to the apparent paradox lies in the fact that strictly speaking, all kinetic terms, induced from the bulk are non-local -- they can be thought of as arising from integrating out a continuum of particles, which includes an infinite number of states, lighter than $\Lambda_3$.~Nevertheless, the couplings of the continuum states to external fields depend on their 4D mass; in particular, light states couple very weakly to the brane fields, making it possible to make sense of `integrating them out', which results in local physics at energies below $L^{-1}$.~This is analogous to how effective 4D physics emerges on the Randall-Sundrum II (RS II) brane \cite{Randall:1999vf}.

\section{Raising the UV cutoff of massive gravity: a detailed account}
\label{sec3}
Having presented a summary of the proposed model, in this section we wish to turn to a more thorough discussion of its dynamics.~We will start with a detailed description of the brane and the bulk theories, quickly overviewed in the previous section.~This will be followed by a  discussion of the procedure by means of which one can effectively `integrate out' the extra dimension, arriving at a weakly-coupled 4D theory of massive gravity. 

\subsection{The brane action}
\label{drgt}
Let us first specify the brane action.~This is given by dRGT massive gravity --  the unique unitary and local non-linear extension of the Fierz-Pauli theory \cite{deRham:2010ik, deRham:2010kj}.~In its diffeomorphism-invariant formulation, in addition to the 4D metric $\gamma_{\mu\nu}$, the theory features 4 auxiliary scalar fields $\phi^p$, with their flavor index running over $p = 0,\dots,3$.~Explicitly, the action reads
\be
\label{sbdy}
\begin{split}
S_{\rm brane} &=\frac{M_4^{2}}{2}\int d^4 x\,\sqrt{-\gamma}\, \bigg[R_4 -\frac{m^2}{4} \sum_{n=2}^4\alpha_n U_n(\mathcal{K}) - 2\Lambda_4\bigg]~,
\end{split}
\ee
where $R_4$ and $\Lambda_4$ are the 4D Ricci scalar and cosmological constant respectively, $\alpha_n$ are constant parameters, $m$ is the graviton's mass (which fixes $\alpha_2 = 2$ in four dimensions) and $M_4$ is the `bare' 4D Planck mass.~The \textit{full} effective 4D Planck scale that we will refer to as $\mpl$ in this section, will receive an extra contribution from bulk dynamics.~Furthermore, the mass/potential terms $U_n$ can be written with the help of the 4D totally antisymmetric symbol $\epsilon$ as follows\footnote{The term linear in $\mathcal{K}^\mu_\nu$ (corresponding to $n=1$) leads to a tadpole on the Minkowski background and thus obstructs having a 4D Poincar\'e-invariant vacuum. We will discard it in the rest of this paper.} 
\be
\label{Vs}
U_n =\epsilon_{\mu_1 \dots \mu_n \mu_{n+1}\dots \mu_4} \epsilon^{\nu_1 \dots \nu_n \nu_{n+1}\dots \nu_4}~\mathcal{K}^{\mu_1}_{\nu_1}\dots \mathcal{K}^{\mu_n}_{\nu_n}~\delta^{\mu_{n+1}}_{\nu_{n+1}}\dots \delta^{\mu_4}_{\nu_4}\equiv \epsilon \epsilon\, \mathcal{K}^n\,,
\ee
where the matrix $\mathcal{K}$ is defined in terms of the auxiliary scalars and the metric in the following way
\be
\label{K}
\mathcal{K}^\mu_\nu = \delta^\mu_\nu- \(\gamma^{\mu\alpha}f_{\alpha\nu}\)^{1/2}\,,\qquad f_{\mu\nu} = \p_\mu\phi^p \p_\nu\phi^q\, \eta_{pq}\,.
\ee
$f_{\mu\nu}$ is a flat auxiliary metric, related to Minkowski by a coordinate transformation. (One can further generalize the theory by defining it with a \textit{curved} $f_{\mu\nu}$, or even by promoting $f_{\mu\nu}$ to a full-fledged dynamical tensor field, which would define a theory of bigravity \cite{Hassan:2011zd}.)~The second equality in \eqref{Vs} defines notational shortcut, which we will often use in the rest of this paper.~We will assume that the boundary metric is coupled minimally to 4D matter, as it is in general relativity.~Finally, we note that the action \eqref{sbdy} is invariant under \textit{internal} $SO(3,1)$ rotations, acting on the auxiliary scalars' flavor index.

\vskip 0.15cm
\noindent
The dynamical equations that follow from varying the action \eqref{sbdy} admit a flat-space solution with the following expectation values
\be
\label{scalavevs}
\langle \gamma_{\mu\nu} \rangle = \eta_{\mu\nu}, \qquad \langle \phi^p \rangle = \delta^p_\mu x^\mu.
\ee
On this background, the scalars' internal indices mix with the spacetime ones, and we will sometimes not make distinction between the two.~One can use diffeomorphism invariance of the theory to fix \textit{unitary gauge}, in which the four scalars are frozen to their background values, $\phi^\mu  = x^\mu$.~In this gauge, \eqref{sbdy} describes a Lorentz-invariant theory of the metric alone.

\vskip 0.15cm
\noindent
Away from unitary gauge and at sufficiently high energies, the most interesting dynamics of massive gravity feature the helicity-2 ($h_{\mu\nu}$) and helicity-0 ($\pi$) polarizations, defined respectively by the following equations 
\be
\label{stu}
\gamma_{\mu\nu} = \eta_{\mu\nu} + h_{\mu\nu}\,,\qquad \phi^p = \delta^p_\mu \( x^\mu +  \eta^{\mu\nu} v_\nu\) \,\qquad v_\mu = a_\mu - \p_\mu\pi\,
\ee
(the helicity-1 mode, on the other hand,  is captured by the Lorentz vector $a_\mu$).~The high-energy limit of interest is then defined as a double scaling limit
\be
\label{declim}
\mpl \to \infty~, \qquad m\to 0~, \qquad \Lambda_3 = \text{finite}\,,
\ee
in which the relevant part of the action (excluding the helicity-1 mode) becomes
\be
\label{dlaction}
\mathcal{L}_{\rm dl} = -\frac{M_4^2}{4}\, h^{\mu\nu} (\hat{\mathcal{E}} h)_{\mu\nu} - \frac{M_4^2\, m^2}{4}\bigg[ \epsilon\epsilon h \p^2 \pi +\beta_2 \epsilon\epsilon h (\p^2 \pi)^2 + \beta_3 \epsilon\epsilon h (\p^2 \pi)^3\bigg] + h^{\mu\nu} T_{\mu\nu}\,.
\ee
Here we have defined $\beta_2 = (3\alpha_3 + 4)/4$ and $\beta_3 = (\alpha_3 + 4\alpha_4)/4$, and used the simplified notation, given in eq.~\eqref{Vs}.

\vskip 0.15cm
\noindent
As remarked multiple times above, $\pi$ has no `independent' dynamics: it only receives its kinetic term through mixing with the helicity-2 polarization of the massive graviton.~This kinetic term is `small' in the sense discussed in the previous section, leading to the low strong coupling scale of the theory, as well as an order-one coupling of $\pi$ to matter (in gravitational units).~To avoid these problems, it would be tempting to try giving $\pi$ an `independent' kinetic term by means of modifying the action for the Nambu-Goldstone sector $\phi^p$ of the theory.~However, one can show that any such (local) modification would clash with unitarity: the dynamics of the Nambu-Goldstone sector of massive gravity is uniquely determined by locality and the absence of extra, pathological degrees of freedom. 

\subsection{Higgs mechanism on AdS$_5$: kinematics}
\label{higgsonads}
One way out could be to give up locality.~A \textit{generic} non-local modification of the theory would put it on shaky grounds, as it would likely reintroduce problems with the basic principles of quantum field theory such as unitarity, causality, etc.~Nevertheless, we know of at least one way to make a 4D field theory non-local without spoiling consistency: embed it in a certain local higher-dimensional spacetime.

\vskip 0.15cm
\noindent
We'd like the higher-dimensional theory to describe AdS gravity in Higgs phase, as outlined in the introductory section.~In order to better understand how the Higgs phenomenon works in this case, we will start with discussing the representation theory of the global symmetry group of 5D AdS: $SO(2,4)$.

\vskip 0.15cm
\noindent
The irreducible, unitary representations of $SO(2,4)$ are labeled by the eigenvalues of the maximal compact subgroup $SO(2) \otimes SO(4) \cong SO(2) \otimes SU(2)_+ \otimes SU(2)_-$, denoted respectively by $E$, $s_+$ and $s_-$.~In particular, $E$ measures energy of a particle in units of AdS curvature $L^{-1}$, which is different from the particle's Lagrangian mass $\bar m$.~The precise relation between the two is (see e.g.~\cite{deWit:2002vz} and references therein)
\begin{align}
\bar m^2 L^2 &= E(E - 4) \quad &s = 0\,,\nonumber \\
\bar m^2 L^2 &= (E+ s - 2)(E - s - 2)\,, \quad &s \ge \,1,\nn
\end{align}
where we have denoted the spin of the particle in question by $s$.~An integer spin-$s$ state with energy $E$ forms an irreducible representation that we will refer to as $D\(E, s/2, s/2\)$.~Moreover, unitarity requires that energy be bounded from below: $E \ge s+2$, and this inequality is only saturated for massless particles.~In terms of the $SO(2,4)$ representation theory, the Higgs mechanism can be understood as the following statement: in the massless limit $E \rightarrow s+2$, a spin-$s$ representation $D\(E, s/2, s/2\)$ becomes \textit{reducilble}, splitting into the following direct sum (see refs.~\cite{Porrati:2001db, Porrati:2003sa, deWit:2002vz, Nicolai:1984hb, heidenreich1981tensor} and references therein):
\begin{align}
D\(E, s/2, s/2\) \overset{E \rightarrow s+2}\longrightarrow D\(s+2, s/2, s/2\) \oplus D\(s+3, (s-1)/2, (s-1)/2\).
\end{align}
The first representation on the right hand side is the massless spin $s$ particle, which `eats up' a \textit{massive} NG boson of spin $s-1$ (the second term in the direct sum) and becomes massive. For the case of a spin-2 graviton in 5D, the NG boson is a \textit{vector} of energy $E = 5$, corresponding to the Lagrangian mass $m_V^2 L^2 = 8$.~The action that describes such a particle is precisely the one in eq.~\eqref{VonAds}, discussed in the context of the \stu formulation of the theory in the introductory section.
\vskip 0.15cm
\noindent
The above discussion has only concerned the linearized limit of massive gravity.~Can the gravitational Higgs mechanism be embedded into a full-fledged nonlinear theory on anti de Sitter space?~To answer this question, we need to first understand the origin of the NG vector $D\(5,1/2,1/2\)$ in such a non-linear theory.~In the scenario of interest, $D\(5,1/2,1/2\)$ will arise as part of a two-particle Hilbert space, formed by a direct product of spin-0 representations, $\mathcal{H}_2 = D(E_+, 0, 0) \otimes D(E_-, 0, 0)$ \cite{Porrati:2001db}.~For concreteness, we will consider the case that the two representations in the product both stem from a conformally coupled AdS scalar.~Such scalars can be quantized in two different ways on anti de Sitter space (depending on the specific boundary conditions one imposes at the spacetime boundary), corresponding to $E_{\pm} = (d \pm 1)/2$ (so that $E_+ + E_- = d$)~\cite{Breitenlohner:1982jf}.~In the case of AdS$_4$, the explicit expression for $\mathcal{H}_2$ can be found e.g.~in ref.~\cite{Porrati:2001db}, and it is straightforward to generalize the formula to AdS$_5$ \cite{heidenreich1981tensor}:
\begin{align} \label{tensordecomp1}
D(E_+, 0, 0) \otimes D(E_-, 0, 0) = \sum_{n = 0}^{\infty} \sum_{s = 0}^{\infty} D\(E_+ + E_- + s + 2 n, s/2, s/2\).
\end{align}
One can see that the NG vector $D\(5,1/2,1/2\)$ does indeed appear in the two-particle Hilbert space of scalar representations of $SO(2,4)$ in $d+1 = 5$.~Importantly, $E_+$ and $E_-$ necessarily have to be \textit{different} for the mechanism to work: had we chosen the same scalar representations on the left hand side of~\eqref{tensordecomp1}, $D\(5,1/2,1/2\)$ would be eliminated from the two-particle Hilbert space by Bose statistics \cite{Porrati:2001db}.~Denoting the two scalars with energy $E_+$ and $E_-$ respectively by $\phi_1$ and $\phi_2$ and taking into account  parity of the composite NG vector $V$ with respect to the interchange $\phi_1 \leftrightarrow \phi_2$\,, we have the following relation between $V$ and the two constituent scalars~\cite{Porrati:2001db}
\be
\label{V}
V = \phi_1 \nabla \phi_2 - \phi_2  \nabla \phi_1\,,
\ee
Moreover, the CFT dual of $V$ is expressed in terms of the CFT duals $\mathcal{O}_{1,2}$ of $\phi_{1,2}$ as
\be
\mathcal{O}_V =  \mathcal{O}_1\p \mathcal{O}_2 - \mathcal{O}_2\p \mathcal{O}_1\,.
\ee
\noindent
The NG vector has precisely the right $SO(2,4)$ quantum numbers to mix with the graviton and become `eaten up' in the Higgs phase of the theory.~Moreover, $D\(5,1/2,1/2\)$ is the only such state: it is easy to convince oneself that no other particle in $\mathcal{H}_2$ can have a linear mixing with the graviton.~Indeed, AdS$_5$ particles can mix at the level of the quadratic action only if their $SO(2,4)$ quadratic Casimirs $C_2$ are identical.\footnote{The quadratic Casimir enters the dynamical equation for a spin-$s$ field on $AdS$ in the following way: $(\Delta - C_2) \Psi_{a_1\dots a_s} = 0$, where $\Delta$ is the Lichnerowicz operator which commutes with covariant derivatives and traces and reduces to $-\Box$ in the flat-space limit of the theory.}~Recalling the explicit form of the quadratic Casimir (see e.g. \cite{deWit:2002vz, Nicolai:1984hb} and references therein)
\begin{align}
C_2 = E(E - 4) + s (s + 2)\,,
\end{align}
one can redily check that no state in $\mathcal{H}_2$ apart from $D(5,1/2,1/2)$ has the value of $C_2$ identical to that of the graviton (which is in the $D(4,1,1)$ representation of $SO(2,4)$ and thus has $C_2 = 8$).~All states on the right hand side of \eqref{tensordecomp1} except  $D(5,1/2,1/2)$ can thus be thought as the `radial' modes of the Higgs sector.

\vskip 0.15cm
\noindent
As a digression, we note that a similar mechanism of mass generation for the electroweak gauge bosons would operate in the Standard Model of particle physics, had its vacuum been AdS.~The reason is that, as it turns out, any chiral gauge symmetry (such as the $SU(2)_L$ of the SM) is bound to be broken down to a vector subgroup by the AdS-invariant boundary conditions of fermions, leading to mass generation for the W-boson even in the absence of the Higgs condensate \cite{Rattazzi:2009ux}.~In this case, the required NG bosons are provided by the geometric bound states of chiral SM fermions, much in the same way as a bound state of conformal scalars gives rise to the gravitational Higgs mechanism.

\vskip 0.15cm
\noindent
The above-described Higgs mechanism for gravity involves mixing between one-particle  (the massless AdS graviton) and two-particle (the composite vector) states, and therefore the graviton mass will only arise at the 1-loop level.~In order to compute it, one can look at the correction to the graviton self-energy from couplings to the two conformal scalars $D(E_\pm,0,0)$.

\subsection{Graviton mass from AdS$_5$ loops}
\label{bigravity}
In fact, it will prove more interesting to start with a setup involving \textit{two} spin-2 states on $AdS_5$, with different `Planck masses' $M_1$ and $M_2$, each coupled to a free conformal scalar.~(In the end we will go back to the case with one spin-2 particle, taking the limit in which the second decouples.)~This corresponds to having a theory of `bigravity', defined by the following action \cite{Aharony:2006hz}
\begin{align} \label{bulkaction1}
S_{\rm bulk} &= \sum_{i = 1,2} \int d^{d+1} x\,\sqrt{-g_i} \Bigg[\frac{ M^{d-1}_i}{2}\(R_i - 2\Lambda\) - \frac{1}{2}\, g_i^{ab}\partial_a \phi_i\partial_b \phi_i - \frac{d-1}{8\, d}\, \phi_i^2  R_i  \Bigg] \nonumber \\
&-\frac{d - E_+}{2} \int d^d x \sqrt{-\gamma_1} \,\phi_1^2+ \int d^d x \sqrt{-\gamma_2} \(\frac{E_-}{2} \phi_2^2 + \phi_2 n_2^{a} \partial_{a} \phi_2\)\,.
\end{align}
Here $\gamma_{1,2}$ are the induced metrics on the brane and  $n_i^a$ are the unit (outward) normals, satisfying $g_i^{ab}n_{ia}n_{ib} = 1$.~So far this is a theory consisting of two decoupled sectors, and as such it is invariant under two distinct sets of diffeomorphisms, $\rm{diff}_1$ and $\rm{diff}_2$,  corresponding to transforming the two pairs of bulk fields (metric plus scalar) separately.~For a negative cosmological constant $\Lambda$, the dynamical equations of the theory admit a solution with $\langle\phi_i \rangle = 0$, both metrics describing AdS space of curvature $L^2 = -d(d-1)/2\Lambda$ (in order to have a flat boundary, one needs to add compensating tension terms for each of the two metrics at $z=0$ as well, which we leave implicit here).

\vskip 0.15cm
\noindent
Solving the dynamical equation for a conformally coupled scalar on AdS yields the following behaviour near the brane (see Appendix~\ref{appa})
\begin{align}
\phi(z\to 0,x) &\approx (L+z)^{E_+} \beta (x) + (L+z)^{E_-} \(\alpha(x)- L^{E_+ - E_-}\beta(x)\) \nn \\
&=\(E_+ - E_-\)  L^{E_+-1} z\beta(x) + L^{E_-}\alpha(x)+ \mathcal{O}(z^2/L^2)\,.
\end{align}
Such a scalar can be quantized with \textit{two} AdS-invariant (Dirichlet or Neumann) boundary conditions, corresponding to $\{\beta = 0,\, \alpha\neq 0\}$ or $\{\alpha = 0,\, \beta\neq 0\}$.~These quantization rules give rise to the $SO(2,4)$ representations $D(E_+,0,0)$ and $D(E_-,0,0)$ respectively \cite{Breitenlohner:1982jf, Klebanov:1999tb}, and we have added the boundary terms in \eqref{bulkaction1} to enforce just these boundary conditions, with the identification $E_1 = E_+$ and $E_2 = E_-$ \cite{marolf2006boundary}.~Consider now adding a small perturbation to the boundary action:
\begin{align} \label{lambdabdy}
S_{\lambda} = - \lambda (E_+ - E_-) \int d^d x \sqrt{-\gamma_1} (E_- \phi_1+  n_1^{a} \partial_{a} \phi_1) \phi_2\,.
\end{align}
This enforces \textit{mixed} boundary conditions on $\phi_{1}(\phi_2)$ in \eqref{bulkaction1}, corresponding to mostly describing the state with AdS energy $E_+ (E_-)$, but with a small admixture of the state with energy $E_- (E_+)$.~More explicitly, the new boundary term \eqref{lambdabdy} correlates the boundary conditions between the two scalars so that their behaviour near the brane becomes
\begin{align}
\label{boundaryconds}
\phi_1(z\to 0, x) & \approx \(E_+ - E_-\)  L^{E_+-1} z\beta_1(x) + L^{E_-}\alpha_1(x) \,, \nn \\
\phi_2(z\to 0, x) & \approx L^{E_-}\beta_2(x) + \(E_+ - E_-\)  L^{E_+-1} z\alpha_2(x)\,,
\end{align}
where the $\alpha$-coefficient of one scalar is related to the $\beta$-coefficient of the other one as follows
\be
\label{boundaryconds1}
\alpha_{1,2} = \mp \lambda (E_+ - E_-) \beta_{2, 1}\,.
\ee
The motivation for choosing such boundary conditions comes from the gauge/gravity duality, and the small coupling $\lambda$ will have a simple interpretation in terms of the dual CFT description of the theory, as will be discussed in more detail in the next section.\footnote{There is in fact an ambiguity in \eqref{lambdabdy} as to which induced metric one should couple the perturbation to: $\gamma_1$, $\gamma_2$, or both.~At the 1-loop level we will be interested in, this won't matter -- all that matters is the background value of the metric (which is the same for both $\gamma_1$ and $\gamma_2$).~Beyond the 1-loop level, we choose to couple the perturbation term to $\gamma_1$, as this is the field whose `Planck mass' we will eventually send to infinity, effectively decoupling its fluctuations.}

\vskip 0.15cm
\noindent
To procceed, we note that a simple  rotation brings us back to the basis of fields with independent boundary conditions 
\begin{align}
\label{chis1}
\chi_1 &= \frac{1}{\sqrt{1 + \tilde{\lambda}^2}}\, (\phi_1 + \tilde{\lambda} \phi_2)\,, \\
\label{chis2}
\chi_2 &= \frac{1}{\sqrt{1 + \tilde{\lambda}^2}}\, (- \tilde{\lambda} \phi_1 + \phi_2)\,,
\end{align}
where we have defined $\tilde \lambda \equiv \lambda (E_+ - E_-)$.~Moreover, the fields $\chi_{1,2}$ have the right boundary conditions to describe irreducible scalar representations $D(E_+,0,0)$ and $D(E_-,0,0)$ and can therefore be quantized in the AdS-invariant way.~Notice, however, that the AdS-invariant quantization will necessarily break one combination of two diffeomorphisms we started with.~This is because $\chi_1$ and $\chi_2$, being linear combinations of $\phi_1$ and $\phi_2$, \textit{do not} have well-defined transformation properties under the full group $\rm{diff}_1\otimes \rm{diff}_2$, but only under the diagonal combination of the two diffs, which remains unbroken.~This, as we will see, results in mass generation for one combination of the original spin-2 particles. 

\vskip 0.15 cm
\noindent
This mass, denoted below by $\bar m$, has been computed previously by several authors in the AdS/CFT limit, which formally corresponds to moving the brane all the way to the AdS boundary $z'= 0$; the calculation for a single bulk spin-2 state has been done in refs.~\cite{Porrati:2001db, Duff:2004wh}, and has been generalized to the case with two spin-2 particles in~\cite{Aharony:2006hz, Kiritsis:2006hy}.~In the two appendices, we extend these calculations in our `regularized AdS/CFT' setup with the brane located at $z' = L$ in the Poincar\' e patch coordinates \eqref{zprime}.~Our calculation is significantly more involved, but we find that the expression for $\bar m$ is not corrected at the leading order in $L$, compared to the results obtained in the AdS/CFT limit.

\vskip 0.15cm
\noindent
To proceed, we note that in terms of the quantized $\chi$-fields the matrix of 2-point functions is diagonal:
\be
\label{chiprop}
\langle \chi_i \chi_j\rangle = \delta_{i j} \tilde G_{E_i}\,,
\ee
where we have defined $E_{1,2} = E_{\pm}$, and $\tilde G_{E_{1,2}}$ are the scalar AdS propagators, satisfying the appropriate (Dirichlet and Neumann) boundary conditions.~The explicit expressions for these propagators are given in Appendix~\ref{appa}.~Furthermore, as discussed extensively around eq.~\eqref{tensordecomp1}, the Nambu-Goldstone vector $D(5,1/2,1/2)$ that the graviton needs to eat up to become massive is only contained in a tensor product of \textit{different} scalar $SO(2,4)$ representations.~This means, in particular, that non-zero contributions to the 1-loop graviton mass come exclusively from the operators in \eqref{bulkaction1} that involve both fields $\chi_1$ and $\chi_2$.~These operators, as can be inferred from eqs.~\eqref{bulkaction1}, \eqref{chis1} and \eqref{chis2}, all couple to the same combination $ (h_{1} - h_{2})/\sqrt{2}$ of the original spin-2 particles, which will therefore acquire mass at one loop.~The resulting quadratic spin-2 Lagrangian schematically reads \cite{Apolo:2012gg}
\be
\label{quadaction}
\mathcal{L} = \frac{1}{4} (M_1^3 + M_2^3)\, h^{(0)}\hat{\mathcal{E}}h^{(0)} + \frac{M_1^3 M_2^3}{M_1^3 + M_2^3}\, h^{(m)}\hat{\mathcal{E}}h^{(m)}  - \sigma \, h^{(m)} h^{(m)}\,,
\ee
where we have defined
\begin{align}
h^{(0)} &= \sqrt{2}\,\frac{ M_1^3 M_2^3}{M_1^3 + M_2^3} \(\frac{1}{M_2^3}\, h_1 + \frac{1}{M_1^3}\, h_2 \),\\
h^{(m)} &= \frac{1}{\sqrt{2}}\( h_1 -  h_2\)\,.
\end{align}
The last term in \eqref{quadaction} describes the loop-generated Fierz-Pauli mass with $\sigma =  \lambda^2/32\pi L^{5}$, while the first two terms denote the kinetic terms for the massless and the massive combinations of the original fields.~The action \eqref{quadaction} also makes it clear that the massless and the massive spin-2 fields couple with different strengths (i.e. have different `Planck masses').~In particular, in the limit $M_1\to \infty$ the massless spin-2 state decouples from all external fields, while the massive one still has a finite `Newton's constant' of order $M_2^{-3}$~\cite{Apolo:2012gg}.~The decoupled massless spin-2 mostly corresponds to the original field $h_1$, while the massive combination---which we will sometimes refer to as the `graviton' below---is mostly $h_2$, and its mass is of order
\be
\label{mbar}
\bar m^2  = \frac{\sigma}{ M^3_2} = \frac{\lambda^2}{32 \pi M^3_2 L^5}\,. 
\ee
We will be mostly interested in this `single-graviton' limit in the rest of this paper. 

\vskip 0.15cm
\noindent
To close the present discussion, we remark that one can integrate out the two scalars $\phi_1$ and $\phi_2$ from the bulk action \eqref{bulkaction1}, thereby arriving at (two copies of) general relativity, corrected by the Coleman-Weinberg `potential' that depends on the two metrics alone:
\begin{align} \label{bulkaction2}
S_{bulk}\(g_1, g_2\) = \sum_{i = 1,2} \int d^{d+1} x\,\sqrt{-g_i} \Bigg[\frac{ M^{d-1}_i}{2}\(R_i - 2\Lambda\) \Bigg]
+ \frac{i}{2}\int d^d x \log \det \Delta(x,z;x',z')\,.
\end{align}
The matrix $\Delta$, defined as $\Delta(x,z;x',z') = \delta^2 S / \delta \phi_i (x,z)\delta \phi_j (x',z')$, can be formally expressed in terms of the following differential operator on anti de Sitter space
\begin{align}
\begin{split}
\Delta &= \Bigg[ \sqrt{-g_1} \(\Box^{\leftrightarrow}_1 - \frac{d-1}{4 d} R_1\) \delta_{1 i} \delta_{1 j} + \sqrt{-g_2} \(\Box^{\leftrightarrow}_2 - \frac{d-1}{4 d} R_2\) \delta_{2 i} \delta_{2 j} \Bigg] \delta^{d+1} (x, z; x', z')  \\
&- \Bigg[ \sqrt{-\gamma_1} \( E_- + n_1\cdot\nabla^{\leftrightarrow}_1\) \delta_{1 i} \delta_{1 j} - \sqrt{-\gamma_2}\, \(E_- + n_2\cdot\nabla^{\leftrightarrow}_2\) \delta_{2 i} \delta_{2 j}\Bigg]\delta(z) \delta (z') \delta^d (x,x')\nonumber\\
& -\sqrt{-\gamma_1}\,  \lambda (E_+ - E_-)\big(E_- (\delta_{1i}\delta_{2j} + \delta_{1j}\delta_{2i}) + n_1\cdot\nabla_1\, \delta_{1 i} \delta_{2 j}  + n_1\cdot\nabla'_1\, \delta_{1 j} \delta_{2 i}\big)\, \delta(z) \delta (z') \delta^d (x,x')\,,
\end{split}
\end{align}
where $\nabla'_i$ ($\nabla_i$) is the covariant derivative with respect to the metric $g_i$, acting on (un)primed coordinates, $n_i\cdot \nabla_i^{\leftrightarrow}\equiv \frac{1}{2}\(n_i\cdot \nabla_i +n_i\cdot \nabla'_i \)$, and similarly for the covariant Laplacian $\Box^{\leftrightarrow}$.

\vskip 0.15cm
\noindent
The expression for the determinant becomes slightly simpler in the limit $M_1\to\infty$, in which the fluctuations of the tensor $g_1$ decouple.~In this case, this field can be substituted by its background value $\bar g$.

\subsection{The quantum strong coupling of the effective brane theory}
\label{scaleup}
Having specified both the boundary and the bulk gravitational actions, eqs.~\eqref{sbdy} and \eqref{bulkaction1}, we are finally in a position to study the strong coupling phenomenon in the effective 4D theory of massive gravity on the brane.~We will work in the limit of the single massive spin-2 state in the bulk (with the other, massless spin-2 state decoupled, as discussed at the end of the previous subsection).~This state will be referred to as $\bar h_{ab}$ (while the full bulk metric is $g_{ab} = \bar g_{ab} + \bar h_{ab}$).~Moreover, its coupling---the higher-dimensional Planck mass, denoted before by $M_2$---will be renamed into~$M_5$\,.~As we have discussed in the previous subsection, the mass of the bulk graviton, $\bar m \sim \tilde\lambda^2/M^3_5 L^5$, is parametrically smaller than the two other scales in the problem, $M_5$ and $L^{-1}$, which we will assume are only mildly separated from each other: $L^{-1} \lesssim M_5$.~Apart from $\bar h_{ab}$, the 5D AdS theory features a tower of `bound states' of the two scalars $\phi_{1,2}$, that have various spins and Lagrangian mass parameters of order $L^{-1}$ and larger.~Because the underlying `fundamental' theory \eqref{bulkaction1} is weakly coupled at energies and momenta below the 5D Planck scale, the `effective' theory of the massive $\bar h_{ab}$ plus the bound states is as well.

\vskip 0.15cm
\noindent
A short comment on the precise meaning of `the low-energy limit' of the bulk theory is in order.~Such a limit is in fact somewhat subtle on AdS, since at energies/momenta, lower than $L^{-1}$ (which we are assuming is a rather high energy scale in this work), the effects of the background curvature become order-one important.~Perhaps a more intuitive definition of the low-energy regime---which we adopt throughout in this paper---arises from the perspective of a 4D \textit{brane} observer.~Indeed, such an observer lives on \textit{flat} space, and can therefore probe the gravitational interactions by conventional means, e.g. by scattering 4D matter particles and measuring the amplitudes.~As a matter of fact, even from the 4D perspective there is a subtlety, as the gravitational sector of the effective brane theory is strictly speaking non-local, containing, in addition to 4D gravity, a gapless continuum of (Kaluza-Klein) states.~Nevertheless, one can still make sense of the 4D low-energy effective field theory: the extra bulk states are `invisible' to an observer, confined to the brane and working at energies/momenta lower than AdS curvature $L^{-1}$ \cite{Randall:1999vf}.~We will return to this point below.

\vskip 0.15cm
\noindent
Back to the bulk theory.~At the level of the quadratic action, its relevant part is given by the Fierz-Pauli theory of eq.~\eqref{FP1}, where the (composite) \stu vector $V_a$, defined in terms of the constituent scalars in eq.~\eqref{V}, transforms under the linearized 5D diffeomorphisms as 
\be
\label{deltaV}
\delta V_a = \bar \xi_a (x,z)\,.
\ee
At the same time, the helicity-2 field $\bar h_{ab}$ transforms as $\delta \bar h_{ab} = \nabla_a\bar\xi_b + \nabla_b\bar\xi_a$.~We will use some of this gauge freedom to fix the bulk coordinates such that the brane sits  straight at $z=0$ (in other words, the `brane bending mode' is gauged away).~In these coordinates, the bulk metric is related to its boundary counterpart (the induced metric on the  brane) as 
\be
\gamma_{\mu\nu}(x) = \delta^a_\mu \delta^b_\nu\, g_{ab}\(x, 0\)=  g_{\mu\nu}|\,.
\ee
As briefly remarked in the previous section, even with the brane frozen at $z=0$, one can further fix the gauge so that the following conditions hold  
\be
\bar h_{\mu z } = \bar h_{zz} = 0\,.
\ee
This still leaves some residual gauge freedom:~namely, consistently with all previous gauge choices, one can choose a non-trivial parameter $\bar \xi_a = L^2 (z+L)^{-2}\,\delta^\mu_a \,\omega_\mu(x)$, which generates the following transformation of the 5D fields
\be
\label{transforms}
\delta \bar h_{\mu\nu} = \frac{L^2}{(z+L)^2} \,(\p_\mu \omega_\nu + \p_\nu\omega_\mu)\,, \qquad \delta V_\mu =  \frac{L^2}{(z+L)^2}\, \omega_\mu\,.
\ee
At the location of the brane $z=0$, this induces the 4-dimensional gauge transformation of the brane metric $\delta h_{\mu\nu} = \p_\mu\omega_\nu + \p_\nu\omega_\mu\,.$~In addition to $h_{\mu\nu}$, the covariant brane theory \eqref{sbdy} contains four \stu scalars $\phi^\mu = x^\mu + \eta^{\mu\nu} v_\nu$, with $v_\mu$ shifting under the 4D brane diffeomorphisms as~$\delta v_\mu = \omega_\mu (x)$.~At the given (linear) order in fields and gauge paramaters, this shift matches with the transformation \eqref{transforms} of the boundary `image' of the bulk \stu field $\bar V_{\mu}(x,0) = V_\mu |$.~At this order, therefore, we will identify
\be
v_{\mu} = V_\mu |\,,
\ee
which should be understood as part of the definition of our theory.~This linearized relation will be sufficient for our purposes of showing how a quadratic kinetic term for the longitudinal $v_\mu$ (that is, the helicity-0 polarization of the 4D graviton) arises from the bulk dynamics.

\vskip 0.15cm
\noindent
Before we procceed, it is instructive to recall how counting of degrees of freedom works from the point of view of a 4D observer.~A massive 5D graviton propagates 9 degrees of freedom.~At energies \textit{well above} $\bar m$, and in the gauge we are working with, these organize into the 5 degrees of freedom of a general-relativistic (helicity-2) graviton, described by $\bar h_{\mu\nu}$, and four extra (helicity-1 and -0) degrees of freedom that live in the 5D \stu field $V_a$\,.~In the limit $\bar m\to0$, the effective 4-dimensional spectrum of the helicity-2 graviton $\bar h_{\mu\nu}(z,x)$ is well-known from previous work on the RS II model~\cite{Randall:1999vf}.~It consists of a gapless continuum of spin-2 KK modes, plus a special, \textit{localized} zero mode $h_{\mu\nu}(x)$.~In the theory under consideration, a short-distance 4D observer would additionally see the spectrum of states stemming from the higher-dimensional field $V_a$, which we will study in a little more detail in what follows. 

\vskip 0.15cm
\noindent
Now, even with a \textit{non-zero} $\bar m$, the four-dimensional KK spectrum of the higher-dimensional field $\bar h_{\mu\nu}$ forms a \textit{gapless continuum} \cite{Dubovsky:2000am}:
\be
\bar h_{\mu\nu}(x,z) = \int_{0}^{\infty} dm \, \bar h^{ m }_{\mu\nu}(x)\,f_{m}(z)\,,
\ee
in full analogy with the RS II case (although the KK wavefunctions $f_m(z)$ will be slightly distorted near the origin, compared to the $\bar m = 0$ case).~On the other hand, the infrared dynamics of the would-be zero mode, $h_{\mu\nu} =  \int_0^{\infty} dm \, \bar h^{ m }_{\mu\nu}(x) f_{m}(0) $, deviates qualitatively from its RS II counterpart.~In particular, as is evident from eq.~\eqref{transforms}, at energies of order $\bar m$ and lower, $h_{\mu\nu}(x) = \bar h_{\mu\nu}|$  acquires mixing with the \stu vector $v_\mu(x) = V_\mu |$, `eating it up' and turning into a long-lived resonance \cite{Dubovsky:2000am, Gabadadze:2017jom}.~The mass scale of the resonance is set by $\bar m$, and its non-zero, small width (suppressed by extra powers of $\bar m L$ compared to its mass \cite{Dubovsky:2000am}) is due to the possibility of decaying into the continuum of KK modes.~Taking $\bar m$ of order of the current Hubble rate (and recalling that $\bar m \ll L^{-1}$) makes the massive resonance completely stable for all practical purposes. 

\vskip 0.15cm
\noindent
Having understood the nature of $V_\mu |$ as the \stu field enforcing gauge invariance of the 4D effective theory, we procceed to study the bulk dynamics of this field.~As discussed around eq.~\eqref{FP1}, in the  limit $\bar m \to 0$,  $V_a$ is a massive AdS vector:
\be
\label{Vaction}
S_V =M^3_5\bar m^2 \int d^5 x\, \sqrt{\bar g} \(-\frac{1}{4}F^{ab} F_{ab} - \frac{4}{L^{2}}V^a V_a\)\,.
\ee
We will be particularly interested in the longitudinal part of $V_a$, defined as $V_a = V^T_a -\p_a\Pi$ ($\nabla^a V_a^T = 0$).~This is because the bounday image $\pi = \Pi|$ of this field---the helicity-0 component of the 4D graviton---is responsible for potential strong coupling of the 4D brane theory \cite{Gabadadze:2017jom}.~On the other hand, the 4D graviton's helicity-1 component, defined in \eqref{stu}, is related to its bulk counterpart as $a_\mu = V^T_\mu|$.\footnote{We note that the procedure of splitting the bulk field $V_a$ into the covariantly transverse and longitudinal components is not unique \cite{Gabadadze:2017jom}.~Namely, there is a (gauge) reduncancy under $V^T_a\to  V^T_a + \nabla_a S$, $\Pi\to \Pi + S$, where $S$ satisfies $\Box S = 0$ and is therefore solved, with the appropriate boundary conditions, by
\be 
S(z,x) = \frac{(z+L)^2}{L^2}\, \frac{K_2\((z+L)\sqrt{-\Box_4}\)}{K_2\(L\sqrt{-\Box_4}\)}\,s(x)\,.
\ee
The brane \stu fields of eq.~\eqref{stu} are given by $a_\mu = V^T_\mu|$ and $\pi = \Pi|$ (notice that while $V^T_a$ is constrained to be 5D transverse, $a_\mu$ does not have to satisfy any 4D constraint). The above redundancy is then realized on these fields as gauge symmetry under $a_\mu \to a_\mu + \p_\mu s$ and $\pi \to \pi + s$.
}

\vskip 0.15cm
\noindent
Using eq.~\eqref{Vaction}, one finds for the $\Pi$ action:
\be
\label{Piaction}
S_{\Pi} = -\frac{4 M^3_5\bar m^2}{L^2}\int d^5 x\, \sqrt{\bar g} \,(\nabla_a\Pi\,\nabla^a\Pi)\,.
\ee
Varying this action yields the dynamical equation $\Box\Pi = 0$, which is solved, with the decaying boundary conditions at $z\to \infty$, by the following function
\be
\label{Pisol}
\Pi(z,x) = \frac{(z+L)^2}{L^2}\, \frac{K_2\((z+L)\sqrt{-\Box_4}\)}{K_2\(L\sqrt{-\Box_4}\)}\,\pi(x)\,,
\ee
where $\Box_4\equiv \eta^{\mu\nu}\p_\mu\p_\nu$ and $K_{1,2}$ are Macdonald functions.~Plugging the above solution back into \eqref{Piaction} then yields the boundary effective action for the 4D field $\pi(x)$:
\be
\label{spi}
S_\pi = -\frac{4 M^3_5\bar m^2}{L^2}\int d^4x\,\pi(x)\,\sqrt{-\Box_4}\, \frac{K_1\(L \sqrt{-\Box_4}\)}{K_2\(L \sqrt{-\Box_4}\)}\,\pi(x)\,.
\ee
Non-locality of this action is a result of `integrating out' a gapless continuum of KK modes.~This can be deduced, for example, by examining the pole structure of the two-point function of $\pi$.~Nevertheless, at sufficiently low energies corresponding to $L \sqrt{-\Box_4} \ll 1$, the action \eqref{spi} can be approximated by a standard, local kinetic term:
\be
\label{spik}
S_{\pi}^{\rm bdy} = \frac{4 M^3_5\bar m^2}{L}\int d^4x\,\pi \Box_4 \pi\,.
\ee
That this approximation is possible is simply an expression of the well-known fact that a massless AdS scalar \textit{localizes} on a RS II brane, acquiring effectively four-dimensional dynamics at distances greater than $L$~\cite{Randall:1999vf}.~At the same time, the non-locality of the action \eqref{spi} is crucial in that it allows to evade the no-go result, forbidding an independent kinetic term for $\pi$ in a local and unitary theory of massive gravity~\cite{Gabadadze:2017jom}. 

\vskip 0.15cm
\noindent
The remainder of this section closely follows Sec.~\ref{summary}, with the only difference that here we keep some of the formulae more general and slightly expand on the schematic discussion of that section.

\vskip 0.15cm
\noindent
As outlined around eq.~\eqref{linduced}, in addition to the $\pi$ kinetic term \eqref{spik}, the brane-induced action also contains a kinetic term for the helicity-2 field $h_{\mu\nu}$~(we stress again that there is also a kinetic mixing between $h_{\mu\nu}$ and $\pi$, induced from the bulk \cite{Gabadadze:2017jom}, but we omit this term here as it is unimportant for the interesting range of the parameters of the theory.).~Adding the brane-induced terms to the original 4D action \eqref{dlaction} yields
\begin{align}
\label{dlactionfinal}
\mathcal{L}_{\rm brane + bulk} &= -\frac{\mpl^2}{4}\, h^{\mu\nu} (\mathcal{E} h)_{\mu\nu} +\frac{4 M^3_5\bar m^2}{L}\pi\Box_4\pi -\frac{M^2_{4 }m^2}{4} \epsilon\epsilon h \p^2 \pi \nn \\&-  \frac{M_4^2\, m^2}{4}\bigg[\beta_2 \epsilon\epsilon h (\p^2 \pi)^2 + \beta_3 \epsilon\epsilon h (\p^2 \pi)^3\bigg]+ h^{\mu\nu} T_{\mu\nu} \,,
\end{align}
where we have defined the effective 4D Planck mass as 
\be
\mpl^2 = M_4^2 + \frac{1}{2}M^3_5 L\,.
\ee
The canonically normalized fields are related to those appearing in the action \eqref{dlactionfinal} as 
\be
\hat h_{\mu\nu} = \mpl\, h_{\mu\nu}\,, \qquad \hat \pi = (M^3_5 L)^{1/2} \, \frac{\bar m}{L} \,\pi \,.
\ee
With this normalization of the fields, the dimensionless parameter $\alpha$  (defined in eq.~\eqref{piTcouplig}), quantifying the strength of the fifth force mediated by the helicity-0 graviton is
\be
\alpha \sim \frac{M^2_4m^2}{(M_4^2 + M^3_5 L/2)^{1/2} (M^3_5 L)^{1/2}}\, \frac{L}{\bar m} \,.
\ee
As discussed in sec.~\ref{summary}, for any relevant choice of the parameters, $\alpha$ is an extremely small number (typically of order $ m L \ll 1$) that tends to zero as the graviton mass $m$ is sent to zero.~The linearized theory therefore avoids the vDVZ discontinuity.

\vskip 0.15cm
\noindent
One can go further and estimate the suppression scales for the most relevant interactions of the (canonically normalized) scalar longitudinal mode of the graviton.~Without making any assumptions about the relative magnitudes of the parameters at hand, the generalized expressions for the two scales, $M_\star$ and $\Lambda_\star$, defined in eq.~\eqref{lambdadefs} are:
\begin{align}
\label{lambdasgeneral}
M^3_\star &\sim  \frac{\bar m^2}{m^2}\, \frac{\mpl M^3_5L}{M^2_4}\,L^{-2} = \frac{\bar m^2}{m^2}\, \frac{(M_4^2 + M^3_5 L/2)^{1/2} M^3_5L}{M^2_4}\,L^{-2} \nn \\
\Lambda^6_\star &\sim \frac{\bar m^3}{m^2}\, \frac{\mpl (M^3_5 L)^{3/2}}{M^2_4}\, L^{-3} = \frac{\bar m^3}{m^2}\, \frac{(M_4^2 + M^3_5 L/2)^{1/2} (M^3_5 L)^{3/2}}{M^2_4}\, L^{-3} \,.
\end{align}
For the (natural) choice of the parameters $m \sim \bar m $, $M^2_4 \sim M^3_5 L \sim \mpl^2$\,, made in Sec.~\ref{summary}, these reduce to the expressions found in \eqref{lambdas}.

\section{Comments on the holographic interpretation}
\label{cft}
The weakly-coupled gravitational Higgs mechanism on AdS$_5$, and in particular its `bigravity' realization discussed in Sec.~\ref{bigravity}, has an interesting interpretation in terms of the strongly-coupled dual CFT \cite{Porrati:2001gx, Aharony:2006hz, Kiritsis:2006hy}.~To simplify the discussion, we will first consider the AdS/CFT limit corresponding to putting the brane at the AdS horizon, and  later comment on the effects of moving it a finite distance away, as relevant for our construction, described in the previous sections. 

\vskip 0.15cm
\noindent
The precise conjecture is that the $\lambda\to 0$ limit of the `bigravity' theory, specified by the action \eqref{bulkaction1} and the boundary conditions \eqref{boundaryconds} and \eqref{boundaryconds1} is dual to a direct product $\text{CFT}_1\times \text{CFT}_2$ of 4-dimensional non-interacting CFTs.~(Indeed, we have seen that in the $\lambda\to 0$ limit, the boundary conditions \eqref{boundaryconds} and \eqref{boundaryconds1} separate and the bulk theory \eqref{bulkaction1} splits into two non-interacting sectors, acquiring invariance under two separate diffeomorphisms, diff$_1\times$diff$_2$.)~In the given limit, the stress tensors of the constituent CFTs separately obey the conformal Ward identities (that is, are conserved and traceless), which corresponds to having two sets of diffeomorphisms and the associated two massless  gravitons, $h_1$ and $h_2$, propagating in the bulk.

\vskip 0.15cm
\noindent
Consider now deforming the product CFT by a double-trace operator of the form 
\be
\label{dtdef}
\lambda\int d^4x\, \mathcal{O}_{1} \mathcal{O}_{2}\,, 
\ee
where $\mathcal{O}_{1}$ and $\mathcal{O}_{2}$ are primary operators, belonging to $\text{CFT}_1$ and $\text{CFT}_2$ and dual to the bulk scalars $\phi_1$ and $\phi_2$ of Sec.~\ref{bigravity}.~The scaling dimensions of these operators will be assumed to obey $\Delta_1 + \Delta_2 = 4$, so that the deformation \eqref{dtdef} is marginal.~In that case, the deformed CFT is still a CFT, albeit with a single set of unbroken conformal symmetries -- those that belong to the diagonal of $\text{CFT}_1\times \text{CFT}_2$.~As shown in~\cite{Witten:2001ua}, adding a double-trace deformation \eqref{dtdef} to the boundary theory corresponds to imposing precisely the mixed boundary conditions \eqref{boundaryconds1} on the dual scalars (with $\lambda$ of eq.~\eqref{dtdef} identified with $\lambda$ of eq.~\eqref{boundaryconds1}).

\vskip 0.15cm
\noindent

\vskip 0.15cm
\noindent
With non-zero $\lambda$, the stress tensors of CFT$_{1}$ and CFT$_{2}$ are no longer individually conserved. Instead, there is one conserved linear combination $T_{\mu\nu}$---the one corresponding to unbroken overall spacetime translations---dual to the massless graviton in the bulk.~The orthogonal spin-2 operator $\tilde T_{\mu\nu}$, on the other hand, is no longer conserved and generically acquires anomalous dimension, proportional to $\lambda^2$.~This can be interpreted as mass generation for the dual spin-2 field.~On the CFT side of the duality, the calculation of this mass/anomalous dimension of $\tilde T_{\mu\nu}$ has been carried out in ref.~\cite{Aharony:2006hz, Kiritsis:2006hy}, and the result agrees exactly with the expression \eqref{mbar}, obtained on the gravity side.~As remarked in sec.~\ref{bigravity}, the massless bulk state can always be decoupled by sending its `Newton's constant' to zero, which on the CFT side corresponds to sending the number of degrees of freedom (the rank of the gauge group) of CFT$_1$ to infinity \cite{Apolo:2012gg}.~This limit would only leave the (interacting) massive spin-2 particle in the bulk. 

\vskip 0.15cm
\noindent
Let us now turn to the case that the AdS space is cut off by the brane, located small distance $L$ away from the AdS boundary.
From the point of view of a four-dimensional brane observer, the massless AdS$_5$ graviton decomposes into a localized zero mode, representing $4D$ gravity on the brane, and a (gapless) continuum of Kaluza-Klein (KK) modes that in the dual field thery forms part of the brane CFT.~The couplings of the 4D graviton to the CFT degrees of freedom are governed by the 4D Planck scale $\mpl^2 \sim M^3_5 L$, with $M_5$ denoting the Planck scale of the bulk gravitational theory.~As to the \textit{massive} bulk spin-2 field, its 4D spectrum consists of a gapless continuum of KK modes, that host a special, quasi-localized resonance \cite{Dubovsky:2000am}.~All of these modes belong to the CFT sector in the field theory dual, the quasi-localized 4D mode representing a spin-2 resonance, made entirely of the CFT degrees of freedom \cite{Gabadadze:2006jm, Gabadadze:2017jom}.~Furthermore, apart from introducing non-zero couplings to gravity, by having put the brane finite distance away from the (would-be) AdS$_5$ boundary we have imposed a UV cutoff $\sim L^{-1}$ on the dual CFT.~Such a cutoff certainly breaks conformal invariance, but only \textit{softly}.~In other words, the breaking is due to \textit{irrelevant} deformations, which have little effect in the infrared\,.~At least as far as the calculation of the bulk spin-2 mass (or, its dual anomalous dimension of the spin-2 operator $\tilde T_{\mu\nu}$) is concerned, we have checked this assertion directly on the gravitational side of the duality.~Our setup with a brane acting as a UV regulator provides a concrete and calculable realization of the soft cutoff in the dual CFT.~This allows to verify, via an explicit calculation, that moving the brane small distance $L$ away from the AdS boundary indeed has no effect on long-distance phenomena, such as mass generation of the bulk spin-2 particle.~This calculation is outlined in Appendix~\ref{appb}.

\subsection*{Acknowledgements}
The work of GG was supported in part by NSF grant PHY-1620039. DP is supported by  the \textit{Origins of the Universe Program} of the Simons Foundation.

\appendix
\section{Scalar boundary conditions and propagators in cutoff AdS}
\label{appa}
To start with, let us lay out the notation and conventions, used in this and the next appendices.~To simplify expressions, $z$ from now on will denote the standard Poincar\'e patch coordinate, that we referred to as $z'$ in the main text.~Therefore, the bulk metric will be written as 
\be
\bar g_{\mu\nu} = \frac{1}{z^2}\,\eta_{\mu\nu}
\ee
with the brane located at $z= L$ (we will often denote $L\equiv \epsilon$ to emphasize that we are working at distances, much larger than the AdS curvature radius).~Greek indices will refer to general $(d+1)$-dimensional spacetime coordinates.

\vskip 0.15cm
\noindent
For a scalar with mass $m^2 = E (E - d)$ in AdS$_{d+1}$, there are naively two possible scaling dimensions for the dual operator $E_{\pm} = \frac{d}{2} \pm \nu$ where $\nu = \sqrt{\frac{d^2}{4} + m^2}$. These two scaling dimensions correspond to two possible quantization schemes, distinguished by whether one imposes Dirichlet or Neumann boundary conditions on the bulk field at the boundary of AdS. From this formula for the scaling dimensions, we see that the mass must  satisfy the Breitenlohner-Freedman (BF) bound $m^2 \ge -\frac{d^2}{4}$ corresponding to the natural condition that the scaling dimension $E$ of the dual operator must be real. However, the scaling dimension must satisfy an additional condition known as the unitarity bound $E \ge \frac{d}{2} - 1$. This means that whenever $m^2 > -\frac{d^2}{4} + 1$, $E_-$ does not satisfy the unitarity bound and only $E_+$ is possible (equivalently, the solution, corresponding to the $E_-$ boundary condition in AdS is non-normalizable). However, for $-\frac{d^2}{4} \le m^2 \le -\frac{d^2}{4} + 1$, both quantization schemes are possible (the two AdS modes are both noralizable) and we must specify the boundary conditions that pick one of the two possibilities in order to define the quantum theory.
\vskip 0.15cm
\noindent
For a free scalar $\phi_{+ (-)}$ of weight $E_{+ (-)}$, the asymptotic behavior near the boundary $z = \epsilon$ is \cite{Breitenlohner:1982jf, Klebanov:1999tb}
\begin{align}
\phi_+ (z,x) &\approx z^{E_+} \beta_{\epsilon} (x) + z^{E_-} \(\alpha_{\epsilon} (x) - \epsilon^{2 \nu} \beta_{\epsilon} (x)\)\,, \nonumber \\
\phi_- (z,x) &\approx z^{E_-} \(\beta_{\epsilon} (x) - \epsilon^{2 \nu} \alpha_{\epsilon} (x)\)+ z^{E_+} \alpha_{\epsilon} (x)\,.
\end{align}
We say that $\phi_+$ is quantized regularly, and the corresponding  $\alpha$ and $\beta$ coefficients are
\begin{align}
\beta_{\epsilon} (x) &\equiv \lim_{z \rightarrow \epsilon} z^{1 - 2 \nu} \partial_z (z^{\nu - \frac{d}{2}} \phi (x))\,, \nonumber \\
\alpha_{\epsilon} (x) &\equiv \lim_{z \rightarrow \epsilon} (z^{\nu - \frac{d}{2}} \phi (x))\,,
\end{align}
on the other hand, $\phi_-$ is quantized irregularly, and for such fields the following is true
 \begin{align}
 \beta_{\epsilon} (x) &\equiv \lim_{z \rightarrow \epsilon} (z^{\nu - \frac{d}{2}} \phi (x)) \nonumber \\
\alpha_{\epsilon} (x) &\equiv \lim_{z \rightarrow \epsilon} z^{1 - 2 \nu} \partial_z (z^{\nu - \frac{d}{2}} \phi (x))\,.
\end{align}
In both cases, in the absence of sources on the boundary, we impose $\alpha_{\epsilon} = 0$. For regular quantization, this corresponds to Dirichlet boundary condition, while for irregular quantization -- to Neumann boundary condition. 

\vskip 0.15cm
\noindent
The $AdS$ (bulk-to-bulk) propagator for a scalar with energy $E_{+(-)}$ in a cut-off AdS is found by calculating the propagator in the usual way, and imposing the Dirichlet (Neumann) boundary condition at $z = \epsilon$, rather than $z = 0$ (that would be relevant for the case of the full Poincar\'e patch of AdS). For the boundary at $z=0$, the bulk-to-bulk propagator for either quantization is:\footnote{Our normalization of $G_E$ corresponds to $(\Box_{AdS} - m^2)\,G (z,x;z',x')= \delta(z - z')\delta(x - x')/\sqrt{g}$ and reduced to the correct position space Feynmann propagator in the flat-space limit\,.} 
\begin{align}
G_E (z,x;z',x') = \frac{\Gamma(E)}{2^\Delta \pi^{d/2} (2 E - d)\Gamma(E - d/2)}\, (-Z)^{-E}\, F\(\frac{E}{2},\,\frac{E + 1}{2},\, E -\frac{d}{2} + 1,\, Z^{-2}\) \nn
\end{align}
where $F ={}_2F_1$ is the standard hypergeometric function and $Z = - \(z^2 + z'^2 + (x - x')^2\)/(2 z z')$ is an AdS-invariant function of the two points $(z,x^{\nu})$ and $(z',x'^{\nu})$ (related to the geodesic distance between these points as $\mu$ as $Z = -\cosh(\mu/L)$)\,.

\vskip 0.15cm
\noindent
For the case of a cut-off $AdS$, the $E_+$ propagator was first calculated in \cite{Muck:1998rr} and the $E_-$ propagator is easily found by the same method:
\begin{align} \label{ModifiedGreens}
\tilde G_{E_1} (z,x;z',x') &= G_{E_1} (z,x;z',x') + \int \frac{d^d k}{(2 \pi)^d} (z z')^{\frac{d}{2}} e^{- i k \cdot (x-x')} \frac{K_{\nu} (z k) K_{\nu} (z' k) I_{\nu} (\epsilon k)}{K_{\nu} (\epsilon k)}\,. \nonumber \\
\tilde G_{E_2} (z,x;z',x') &= G_{E_2} (z,x;z',x') - \int \frac{d^d k}{(2 \pi)^d} (z z')^{\frac{d}{2}} e^{- i k \cdot (x-x')} \frac{K_{\nu} (z k) K_{\nu} (z' k) I_{1-\nu} (\epsilon k)}{K_{1-\nu} (\epsilon k)}\,,
\end{align}
where $K_{\nu}, I_{\nu}$ are modified Bessel functions of the second kind. 

\vskip 0.15cm
\noindent
Let us now consider two scalars $\phi_{1, 2}$ in $AdS_{d+1}$, both with mass $m^2$ and quantized with `energies' $E_1 = E_+$ and $E_2 = E_-$.~In the fueld theory dual, this corresponds to a pair of non-interacting CFTs, having a scalar operator of dimension $E_+$ and $E_-$ respectively.~Adding a double trace deformation $W = -\lambda\int d^d x\,  \mathcal{O}_1 \mathcal{O}_2$ that couples the two CFTs amounts, from the bulk perspective, to adding precisely the boundary term of eq.~\eqref{lambdabdy} at $z = \epsilon$.~This modifies the boundary conditions, giving rise to a $\alpha_{\epsilon i}$ term in the asymptotic behavior of scalar $\phi_i$ \cite{Klebanov:1999tb, Aharony:2006hz}:
\begin{align} \label{deformation}
\alpha_{\epsilon 1} - \epsilon^{2 \nu} \beta_{\epsilon 1}&= - \lambda (E_1 - E_2) (\beta_{\epsilon 2} - \epsilon^{2 \nu} \alpha_{\epsilon2}) \nonumber \\
\alpha_{\epsilon 2} &= \lambda (E_1 - E_2) \beta_{\epsilon 1}\,.
\end{align}
Nevertheless, a simple  rotation brings us back to the basis of fields with independent boundary conditions: 
\begin{align}
\chi_1 &= \frac{1}{\sqrt{1 + \tilde{\lambda}^2}} (\phi_1 + \tilde{\lambda} \phi_2) \\
\chi_2 &= \frac{1}{\sqrt{1 + \tilde{\lambda}^2}} (- \tilde{\lambda} \phi_1 + \phi_2)\,,
\end{align}
where we have defined $\tilde \lambda \equiv \lambda (E_1 - E_2)$\,. In terms of the $\chi$-fields, the matrix of 2-point functions is diagonal: $\langle \chi_i \chi_j\rangle = \delta_{i j} \tilde{G}_{E_i}$\,, and the two point functions for the original fields $\phi_i$ can be easily solved for.~The off-diagonal two-point function with boundary at $z = \epsilon$ starts linear in the deformation
\begin{align}
\label{nondiag}
\langle \phi_1 \phi_2 \rangle_{\epsilon} &=  \lambda (E_1 - E_2) (\tilde{G}_{E_1} - \tilde{G}_{E_2})\,. \nonumber \\
&= \langle \phi_1 \phi_2 \rangle_{0} + \tilde \lambda \int \frac{d^d k}{(2 \pi)^d} (z z')^{\frac{d}{2}} e^{- i k \cdot (x-x')} K_{\nu} (z k) K_{\nu} (z' k) \Big( \frac{I_{\nu} (\epsilon k)}{K_{\nu} (\epsilon k)} + \frac{I_{1-\nu} (\epsilon k)}{K_{1-\nu} (\epsilon k)}\Big)\,,
\end{align}
where $\langle \phi_1 \phi_2 \rangle_0 \equiv \lambda (E_1 - E_2) (G_{E_1} - G_{E_2})$ is the unperturbed correlation function for boundary at $z = 0$ and we have used eqs.~\eqref{ModifiedGreens}. The diagonal correlators in $\langle\phi_i \phi_j \rangle_{\epsilon}$ differ from the unperturbed Green's functions only at second order in $\lambda$.
\vskip 0.15cm
\noindent
A closed-form expression for the integral in $\epsilon$-dependent part of \eqref{nondiag} is generically unknown; however, for $\nu = 1/2$ corresponding to a conformally coupled scalar, one can rewrite this formula in a more suggestive way.~Indeed, using the definition of the modified Bessel function $K_\nu (u) = \frac{\pi}{2 \text{sin}(\nu u)} (I_{-\nu} (u) - I_{\nu} (u))$, we notice first that the unregularized correlator \eqref{nondiag} can be written as 
\begin{align} \label{diffGreens}
\langle \phi_1 \phi_2 \rangle _0 & = \frac{2 \tilde \lambda}{\pi}\, \sin(\nu \pi) \int \frac{d^d k}{(2 \pi)^d} (z z')^{\frac{d}{2}} e^{-i k \dot (x - x')} K_{\nu} (k z) K_{\nu} (k z')\,.
\end{align}
Focussing now on $\nu = 1/2$, the expressions for the modified Bessel functions simplify considerably
\footnote{In particular,
\begin{align} \label{Bessel}
I_{\frac{1}{2}} (u)= \sqrt{\frac{2}{\pi}} \frac{\sinh (u) }{\sqrt{u}}\,,\qquad 
K_{\frac{1}{2}} (u) &= \sqrt{\frac{\pi}{2}} \frac{e^{-u}}{\sqrt{u}}\,.
\end{align}}, which, after some algebra, allows one to rewrite the non-diagonal two-point function \eqref{nondiag} in the following way
\begin{align}
\langle \phi_1 \phi_2 \rangle_{\epsilon} =  
  \frac{(z z')^{\frac{d-1}{2}}}{[(z-\epsilon) (z' - \epsilon)]^{\frac{d-1}{2}}}\,\langle \phi_1(z-\epsilon, x) \,\phi_2(z'-\epsilon, x') \rangle _0\,.
\end{align}
(The last factor in this expression is given by replacing $z \to z - \epsilon$, $z' \to z' - \epsilon$ and $\nu \to 1/2$ on the right hand side of \eqref{diffGreens}.)
One can now readily expand this expressin in small $\epsilon$:
\begin{align}
\label{nondiagprop}
\langle \phi_1 \phi_2 \rangle_{\epsilon} &= \langle \phi_1 \phi_2 \rangle_0 + \epsilon \Big(\frac{z + z'}{z z'} \Big) \Big[ (Z + 1) \langle \phi_1 \phi_2 \rangle_0' + \frac{d-1}{2} \langle \phi_1 \phi_2 \rangle_0 \Big] \nn \\&\equiv  \langle \phi_1 \phi_2 \rangle_0 + \epsilon \,\Big(\frac{z + z'}{z z'} \Big) F(Z)
\end{align}
where prime denotes differentiation with respect to the $AdS$-invariant $Z$.~As is evident from this equation, the proper dimensionless expansion parameter is $\epsilon (z + z')/(zz')$\,, which is small as far as both of the two coordinates, $z$ and $z'$, are sufficiently far into the bulk: $\epsilon \ll z\,, z'.$

\section{Graviton self-energy at one loop}
\label{appb}

One way to understand the Higgs mechanism in field theory is through the appearance of a massive pole in the propagator of the gauge boson due to mixing with the Goldstone boson(s) at low energies.~For example, for a $U(1)$ gauge boson $A^\mu$, mixing with the Goldstone $\pi$ through an operator $mA^\mu\p_\mu\pi$ gives rise to (a non-local contribution to) the self-energy $\Sigma_{\mu\nu}(p) = (\eta_{\mu\nu} - p_{\mu}p_\nu/p^2)\,\Sigma(p^2)$\,, and $\Sigma(p^2\to 0) \ne 0$ signals mass generation for $A_{\mu}$.\footnote{In the case under consideration, $\Sigma(p^2\to 0) = m^2$, as required by (Abelian) Higgs mechanism.}~(Notice that $\Sigma_{\mu\nu}$ is transverse and mass generation is perfectly consistent with gauge invariance.)~Likewise, Higgs mechanism for gravity on $AdS$ can be grasped by looking at the proper non-local piece in the (position-space) graviton self-energy $\Sigma_{\mu\nu,\alpha\beta}(z, x; z', x')$, arising from the exchange of an intermediate NG vector $V_\mu$.~(In all that follows, indices from the middle of the Greek alphabet, $\mu, \nu, \dots$, will refer to unprimed coordinates while indices from the beginning of the alphabet, $\alpha, \beta, \dots$, will refer to primed coordinates.)~At long distances, this non-local self-energy takes the form 
\be
\label{beta}
\Sigma_{\mu\nu,\alpha\beta} (z, x; z', x')  \overset{\mu \rightarrow \infty}\longrightarrow \beta\, \Pi^{\mu\nu,\alpha\beta} (z, x; z, x')
\ee
where $\Pi^{\mu\nu,\alpha\beta}$ is the properly normalized projector onto transverse, traceless tensors and $\beta = \bar m^2 \ne 0$ signals that the graviton has gained a mass (here $\mu$ denotes the AdS-invariant geodesic distance between the points $(z,x)$ and $(z',x')$). The relevant mixing of the graviton with the vector Goldstone boson has the form $h^{\mu\nu}\(\nabla_\mu V_\nu - \eta_{\mu\nu}\nabla\cdot V\)$ as can been seen from the diffeomorphism-invariant formulation of the  Fierz-Pauli theory \eqref{FP1}.~Integrating out $V_\mu$ in that theory generates the following \textit{gauge-invariant} correction to the graviton action at long distances
\be
S = S_{\rm EH} - \frac{1}{4}\,\mpl^{d-1} \bar m^2\int d^{d+1} x\sqrt{\bar g}\, h^{\mu\nu}\,\Pi_{\mu\nu}^{~~\rho\sigma}\,h_{\rho\sigma}\,.
\ee
Here $S_{\rm EH}$ denotes the Einstein-Hilbert action, linearized around anti de Sitter space, $\bar g_{\mu\nu}$ is the background metric and $h^{tt}_{\mu\nu} = \Pi_{\mu\nu}^{~~\rho\sigma}\,h_{\rho\sigma}$ is the transverse-traceless part of the metric fluctuation -- gauge invariance requires that the action depend on this precise part of $h_{\mu\nu}$.~More explicitly, in $d+1$ dimensions, $h^{tt}_{\mu\nu}$ reads
\begin{align}
\label{htt}
h^{tt}_{\mu\nu} &= h_{\mu\nu} + \frac{1}{\Delta - \frac{4\Lambda}{d-1}}\(D_{\mu} D^{\lambda}h_{\lambda\nu}+D_{\nu} D^{\lambda}h_{\lambda\mu}\) + \frac{d - 1}{\(\Delta - \frac{4\Lambda}{d-1}\)\(d\Delta - \frac{2\Lambda(d+1)}{d-1}\)}\,D_\mu D_\nu D^\rho D^\sigma h_{\rho\sigma}  \nn\\
& - \frac{1}{d\Delta - \frac{2\Lambda(d+1)}{d-1}} \, D_\mu D_\nu h +\frac{\frac{2\Lambda}{d-1} - \Delta}{d\Delta - \frac{2\Lambda(d+1)}{d-1}}\,\bar{g}_{\mu\nu} h - \frac{1}{d\Delta - \frac{2\Lambda(d+1)}{d-1}}\,\bar{g}_{\mu\nu} D^\rho D^\sigma h_{\rho\sigma}\,,
\end{align}
where $\Delta$ denotes the Lichnerowicz operator, acting on tensor, vector and scalar fields as $\Delta h_{\mu\nu} = -\Box h_{\mu\nu} - 2 R_{\mu\rho\nu\sigma} h^{\rho\sigma} + 2 R^\rho_{~(\mu} h_{\nu)\rho}$\,, $\Delta V_\mu = (-\Box + \frac{2\Lambda}{d-1}) V_{\mu}$\,, $\Delta\phi = -\Box\phi$ respectively \cite{lichnerowicz1962}.~With this definition, $\Delta$ commutes with all covariant derivatives and traces and can thus be treated as a number.~The second term in \eqref{htt} gives rise to a pole in the graviton propagator, which stems from an exchange of a spin-1 state satisfying the wave equation $(\Delta - \frac{4\Lambda}{d-1}) V_{\mu} = 0$.~This is precisely the Goldstone vector with mass $m_V^2 = -4\Lambda /(d-1)= 2 d/L^2$.

\vskip 0.15cm
\noindent
The calculation undertaken in this appendix, albeit for full Poincar\'e patch of $AdS_4$, has been done in several previous papers \cite{Porrati:2001db, Porrati:2003sa, Duff:2004wh, Aharony:2006hz, Kiritsis:2006hy}.~Here we will be extending the calculation to the case of \textit{regularized} $AdS_{d+1}$ and computing not only the mass but also the first order correction to the mass (or, rather, the form factor) due to regularization.~

\vskip 0.15cm
\noindent
We will start with a theory of two gravitons coupled to two scalar fields \eqref{bulkaction1} and integrate out the scalars at one loop, which gives rise to the graviton self-energy matrix 
\be
\Sigma^{\mu \nu, \alpha \beta}_{i j} = 8 \pi G \langle T^{\mu \nu}_i (z, x) T^{\alpha \beta}_j (z', x')\rangle,\quad  i, j = 1,2
\ee
where $T^{\mu \nu}_i$ is the stress energy tensor of for the scalar $\phi_i$. Since only the off-diagonal part of the self-energy will induce a mass for one combination of the gravitons (which we denote here by $h_{\mu\nu}$), we need only compute the off-diagonal two point function of stress tensors $\langle T^1_{\mu \nu} (x, z) T^2_{\alpha \beta} (x', z') \rangle$.~The stress tensor of a single conformal scalar given by
\begin{align}
T_{\mu \nu} = b_1 \nabla_{\mu} \phi \nabla_{\nu} \phi + b_2 \phi \nabla_{\mu}  \nabla_{\nu} \phi + b_3 (\partial \phi)^2 + b_4 \bar g_{\mu \nu} \phi^2\,,
\end{align}
where the constants $d_i$ have been defined as follows 
\be
 b_1 = \frac{d+1}{2 d}\,, \quad b_2 = -\frac{d-1}{2 d}, b_3 = -\frac{1}{2 d}\,,\quad  b_1 = -\frac{(d-1)^2}{8 d}  \,.
\ee
Once the relevant part of the self-energy is computed, we will express it in a basis of bitensors $O^i_{\mu \nu, \alpha \beta} (x, z; x', z'), \,i = 1, \dots 11$, invariant under the symmetries of the regularized AdS$_{d+1}$.~In the case of the full (unregularized) AdS spacetime considered in \cite{Porrati:2001db, Porrati:2003sa, Duff:2004wh, Aharony:2006hz, Kiritsis:2006hy}, the  2-point function is maximally symmetric (invariant under the full $SO(2, d)$ isometry group) and correspondingly can be expressed in terms of a symmetric subset of the above bitensor basis.~The basis of symmetric bitensors can be written in terms of the following elementary (bi-)tensors \cite{allen1986vector}
\begin{align}
\bar g_{\mu \nu} (x, z) \,\text{ (or } \bar g_{\alpha \beta} (x', z') ) \qquad N_{\mu} (x, z; x', z')\, \text{ (or } N_{\alpha} (x', z')) \qquad \hat{G}_{\mu \alpha} (x, z; x', z')\,,
\end{align}
where $\bar g_{\mu \nu}$ denotes the background metric, $N_{\mu}$ is the unit tangent vector at point $(x, z)$, pointing along a geodesic from $(x, z)$ towards $(x', z')$, and $\hat{G}_{\mu \alpha}$ is closely related to the parallel propagator $\hat{g}^\mu_\alpha$.~(When contracted with a vector $V_{\mu}$ at $(x, z)$, the latter object gives the parallel-propagated vector $\hat{g}^\mu_\alpha\,  V_{\mu} = V_{\alpha}$ along the geodesic.)~The exact relation between $\hat G$ and $\hat g$ is \cite{Duff:2004wh}
\begin{align}
\hat{G}_{\mu \alpha} = \hat{g}_{\mu \alpha} + (Z + 1) N_{\mu} N_{\alpha}\,.
\end{align}
The 5 maximally symmetric bitensors are \cite{allen1987evaluation}:
\begin{align}
O_1 &= \bar g_{\mu \nu} \bar g_{\alpha \beta} \,,\nonumber \\
O_2 &= N_{\mu} N_{\nu} N_{\alpha} N_{\beta}\,, \nonumber \\
O_3 &=  \hat{G}_{\mu \alpha} \hat{G}_{\nu \beta} + \hat{G}_{\mu \beta} \hat{G}_{\nu \alpha}\,,  \nonumber \\
O_4 &= \bar g_{\mu \nu} N_{\alpha} N_{\beta} + \bar g_{\alpha \beta} N_{\mu} N_{\nu} \,,\nonumber \\
O_5 &= \hat{G}_{\mu \alpha} N_{\nu} N_{\beta} + \hat{G}_{\nu \beta} N_{\mu} N_{\alpha} + \hat{G}_{\mu \beta} N_{\nu} N_{\alpha} + \hat{G}_{\nu \alpha} N_{\mu} N_{\beta}\,.
\end{align}
When AdS is cut off by a boundary at $z= \epsilon$, there is another elementary tensor we can add to our building blocks---the outward unit normal vector along the $z$-direction $n_{\mu} (x, z)$---from which new bitensors can be constructed. With this, there are 6 other bitensors, $O_6\,, \dots, O_{11}$, one must add to our basis in a regularized AdS.~These bitensors will not play a role in computing the correction to the graviton mass/form factor and we will therefore keep them implicit throughout the calculation.

\vskip 0.15cm
\noindent
To extract the graviton mass, we need to evaluate the coefficient $\beta$ in eq.~\eqref{beta}.~Denoting by $H$ the regularized 2-point function $\langle \phi_1 \phi_2 \rangle_{\epsilon}$ computed in \eqref{nondiagprop}, one can find the 2-point function of stress tensors by applying Wick contractions on the elementary scalars
\begin{align}
\label{2ptfT}
\langle T^1_{\mu \nu} (z, x) T^2_{\alpha \beta} (z', x') \rangle &= b_1^2 \Big[ \nabla_{\mu} \nabla_{\alpha} H \,\nabla_{\nu} \nabla_{\beta} H + \nabla_{\mu} \nabla_{\beta} H \, \nabla_{\nu} \nabla_{\alpha} H\Big] + \dots \,.
\end{align}
We'd like to expand this expression in terms of  our generalized bitensor basis $O_i\,,\,i = 1, \dots 11\,,$ for cutoff AdS.~In doing so, the following formulas prove useful (most of which can be found in table~1 of \cite{allen1986vector}):
\begin{align}
\begin{split}
\nabla_{\mu} Z &= \sqrt{Z^2 - 1} N_{\mu}\,,  \\
\nabla_{\mu} z &= z n_{\mu}  \,,\\
\nabla_{\mu} \sqrt{Z^2 - 1} &= Z N_{\mu}\,,  \\
\nabla_{\mu} N_{\alpha} &= \frac{\hat{G}_{\mu \alpha} - Z N_{\mu} N_{\alpha}}{\sqrt{Z^2 - 1}}\,,  \\
\nabla_{\mu} N_{\nu} &= \frac{Z}{\sqrt{Z^2-1}} ( \bar g_{\mu \nu} - N_{\mu} N_{\nu}) \,, \\
\nabla_{\mu} n_{\nu} &= n_{\mu} n_{\nu} - \bar g_{\mu \nu}  \,,\\
\nabla_{\mu} \hat{G}_{\nu \alpha} &= \sqrt{Z^2 - 1}\, \bar  g_{\mu \nu} N_{\alpha} \,. \\
\end{split}
\end{align}
Besides derivatives, we will also need the expressions for the following contractions of these (bi-) tensors:
\begin{align}
\begin{split}
 \bar g^{\mu \nu} \hat{G}_{\mu \alpha} N_{\nu} &= Z N_{\alpha} \,, \\
 \bar g^{\mu \nu} \hat{G}_{\mu \alpha} \hat{G}_{\nu \beta} &= \bar g_{\alpha \beta} + (Z^2 - 1) N_{\alpha} N_{\beta} \,, \\
\bar g^{\mu \nu} N_{\mu} n_{\nu} &= \frac{-1}{\sqrt{Z^2 - 1}} \Big( \frac{z}{z'} + Z \Big) \,, \\
 \bar g^{\mu \nu} \hat{G}_{\mu \alpha} n_{\nu} &= \frac{z}{z'} n_{\alpha} -\sqrt{Z^2-1} N_{\alpha}\,. \\
\end{split}
\end{align}
Using these relations, one can expand the 2-point function of the stress tensors \eqref{2ptfT} into the basis of 11 bitensors, discussed above.~The ultimate goal is to extract the part of the graviton self-energy that's due to the exchange of the massive NG vector, `eaten up' in the Higgs phase of the theory. This is given by the following expression (appropriately symmetrized):
\begin{align}
\label{Pispin-1}
\Pi^{\text{spin-1}}_{\mu \nu \alpha \beta} = -2 \nabla_{\beta} \nabla_{\nu} D_{\mu \alpha} + \text{permutations}
\end{align}
where $D_{\mu \alpha}$ is the two point function for a massive vector field with mass $m^2 = 2 d/L^2$ on AdS$_5$ \cite{allen1986vector}.~At the zeroth order in the deformation $\epsilon$, both  the graviton self-energy $\Sigma$ and the spin-1 propagator $D_{\mu \alpha}$ are the function of the AdS-invariant quantity $Z$ alone.~Putting the brane at $z = \epsilon$ breaks the isometries of AdS and introduces dependence on $z$ and $z'$ separately, so that the total self-energy can be written to $\mathcal{O}(\epsilon)$ as $\Sigma_\epsilon = \(1 + \epsilon q(z,z')\) \cdot \Sigma_0 + \Sigma^{(1)}_\epsilon$.~Here $\Sigma_0$ is expressable through the AdS-invariant bitensors $O_1\,,\dots O_5$\,, while $\Sigma^{(1)}_\epsilon$ (a combination of the bitensors $O_6$ through $O_{11}$) is the part whose tensor structure breaks AdS isometries -- the precise form of the latter quantity won't matter for our purposes.~The~$\mathcal{O}(\epsilon)$~correction to the graviton form-factor we're after is given by $\epsilon q(z,z')$, which is in general a $z$ and $z'$-dependent function (as directly follows from dimensional analysis).~With this in mind, we obtain:
\begin{align} \label{SET2point}
\begin{split}
\langle T^1_{\mu \nu} T^2_{\alpha \beta} \rangle &= O_1 \Bigg[\frac{1}{8 d^2} \Big( (d-1) H + 2 Z H' \Big)^2 - \frac{d^2 +4 d + 1}{4 d^2} H'^2 + \frac{(d-1)^2}{4 d^2} H H''\bigg]  \\ 
&+ \bar{\epsilon}\, O_1 \bigg [GF \frac{(d-1)^3}{8 d^2} +G F'\frac{(d-1)^2}{4 d^2} Z + G' F \(\frac{(d-1)(d-3)}{4 d^2} Z - \frac{d-1}{2 d^2} \)  \\
&+ G' F' \Big( \frac{d+1}{2 d^2} Z^2 + \frac{5-d^2}{4 d^2} Z - \frac{(d-1)(d+3)}{4 d^2} \Big)  \\
&+ G'' F \frac{d-1}{2 d^2} Z(Z+1) + G''F' \frac{1}{d^2} (Z+1)\bigg ]   \\
&+ O_2 (Z^2-1)^2\bigg [\frac{3 d^2 + 2 d + 3}{4 d^2} H''^2 - \frac{d^2 - 1}{d^2} H' H''' + \frac{(d-1)^2}{4 d^2} H H''''\bigg ]  \\
&+ O_3 \bigg [\frac{(d+1)^2}{4 d^2} H'^2 + \frac{(d-1)^2}{4 d^2} H H''\bigg]  \\
&+ O_4 (Z^2-1) \bigg [-\frac{d^3 + 3 d^2 + d - 1}{4 d^2} H'^2 + \frac{2 (d-1)(d^2 + 2 d - 1)}{8 d^2} H H''  \\
&+ \frac{(d-1)^2}{4 d^2} Z H H''' - \frac{(d+3)(d-1)}{4 d^2} Z H' H''\bigg ]  \\
&+\bar{\epsilon}\, O_4 (Z^2-1) \bigg [-G' G''\,\frac{d^2+2 d + 5}{4 d^2} - G''^2 \frac{(d+1)^2}{d^2} (Z+1) + G G''' \frac{(d-1)^2}{4 d^2}  \\
&+ G' G''' \frac{(d-1)}{d^2} (Z+1) \bigg ] \\
&+ O_5 (Z^2-1) \bigg[\frac{(d-1)^2}{4 d^2} H H''' - \frac{d^2 - 2 d - 3}{4 d^2} H' H''\bigg ]  + \dots \,,
\end{split}
\end{align}
where the ellipses deonte the part of the 2-point function whose tesor structure breaks AdS isometries (that is, depends on the bitensors $O_6\,\dots O_{11}$).~Furthermore, here (and from now on), a prime will denote a partial derivative with respect to $Z$ and $\bar{\epsilon} \equiv \epsilon \Big( z^{-1} + z^{'-1}\Big)$. In the $\epsilon \rightarrow 0$ limit, this expression reduces to the correct two-point function for the stress tensors in the full Poincar\'e patch of AdS, and for $d = 3$ reproduces the formula, found in \cite{Duff:2004wh}. 

\vskip 0.15cm
\noindent
As remarked around eqs.~\eqref{beta} and \eqref{Pispin-1}, to extract the graviton mass we need to look at the large-distance (that is, large $\vert Z \vert$) behavior of \eqref{SET2point} and match it to the large $\vert Z \vert$ limit of the (twice-differentiated) propagator of the massive NG vector. The coefficient of proportionality can then be identified as $\bar m^2$ \cite{Porrati:2001db, Porrati:2003sa, Duff:2004wh}.

\vskip 0.15cm
\noindent
Let us slightly change our basis of AdS-invariant bitensors by introducing the following three traceless combinations (essentially a $(d+1)$-dimensional generalization of the basis of ref. \cite{Duff:2004wh})
\begin{align}
T_1 &= \frac{1}{d (d Z^2 + 1)} \Big( O_1 + (d+1)^2 O_2 - (d+1) O_4 \Big)\,, \nonumber \\
T_2 &= -\frac{1}{d} O_1 + (Z^2 - \frac{1}{d}) O_2 + \frac{1}{2} O_3 + \frac{1}{d} O_4 - \frac{1}{2} Z O_5 \,,\nonumber \\
T_3 &=  \frac{1}{2 Z} (-4 Z O_2 + O_5) \,.
\end{align}
Out of these three bitensors, one can construct a family of transverse traceless bitensors, parametrized by a single function $a(Z)$.~Generalizing again from the calculation in $d +1 = 4$ of \cite{Duff:2004wh}, we have:
\begin{align}
\mathcal{T}[a] = &a (d \,Z^2 +1) T_1 + \Big[ \frac{d}{(d+2)(d - 1)} (Z^2 - 1)^2 a'' + d Z (Z^2 - 1) a' + \frac{d+1}{d - 1} (d Z^2 - 1) a \Big] T_2 \nonumber \\
+ &\Big[ (Z^2 - 1) Z a' + (d + 1) Z^2 a\Big] T_3\,.
\end{align}
Choosing $a_n (Z) = Z^{-n}$, one generates the family of transverse-traceless bitensors $\mathcal{T}_n$:
\begin{align}
\mathcal{T}_n = &\frac{d\cdot Z^2 +1}{Z^n} T_1 + \Big[ \frac{d \cdot n (n+1)}{(d+2)(d - 1)} \frac{(Z^2 - 1)^2}{Z^{n+2}} -n \cdot d \frac{Z^2 - 1}{Z^n} + \frac{d+1}{d - 1} \frac{d\cdot  Z^2 - 1}{Z^n} \Big] T_2 \nonumber \\
+ &\Big[ -n \frac{Z^2 - 1}{Z^n} + (d + 1) \frac{1}{Z^{n-2}}\Big] T_3\,.
\end{align}
The 2-point function of the stress tensors \eqref{SET2point} can be decomposed into this transverse, traceless basis (e.g. using Mathematica), and the result reads 
\begin{align} \label{SET2ptTT}
\langle T^1 T^2\rangle = c_{d+2} \mathcal{T}_{d+2} + c_{d+3} \mathcal{T}_{d+3} + \dots \,.
\end{align}
The coefficients can be expanded in powers of the `UV cutoff': $c_i = c_i^{(0)} + \epsilon c_i^{(1)} + \mathcal{O}(\epsilon^2)$, and the zeroth order terms in $\epsilon$ give just what the 2-point function reduces to in the full AdS limit.~The same expandion of the twice-differentiated propagator \eqref{Pispin-1} of a $(4+1)$-dimensional AdS vector of mass $m^2_V = 8 L^{-1}$ yields
\begin{align}
\label{spin-1prop}
\Pi^{\text{spin-1}} = -\frac{45}{16 \pi^2} \mathcal{T}_6 + \dots\,,
\end{align}
where ellipses denote terms that are sub-leading in the $|Z|\to \infty$ limit.~The graviton mass $\bar m^2$ can then be extracted by calculating the coefficient $c_6$ in \eqref{SET2ptTT} and comparing to eqs.~\eqref{beta}, \eqref{Pispin-1} and \eqref{spin-1prop}
\begin{align}
\label{msq}
\bar m^2 = \frac{ \lambda^2}{32  \pi M_5^3 L^5}\bigg[ 1 + \frac{4 \epsilon}{5} \( \frac{1}{z} + \frac{1}{z'}\)\bigg ]\,.
\end{align}
The $\mathcal{O}(0)$ part is the induced mass of the graviton, which agrees with the result found in \cite{Aharony:2006hz}, while the $\mathcal{O}(\epsilon)$ part is the form factor correction to the spacetime being cut off at $z = \epsilon$.~In the long-distance limit, $\epsilon \ll z, z'$, this correction is necessarily small compared to the mass, which dominates the infrared limit of the graviton's form-factor; it only becomes important once energy scales ($\mu = z^{-1}$) close to the cutoff $\mu \sim L^{-1}$ are probed. 

\vskip 0.15cm
\noindent
We note that the functional form of the result \eqref{msq} could have been guessed based on dimensional analysis alone.

\bibliographystyle{utphys}
\bibliography{bibliography}

\end{document}